\documentclass[useAMS,usenatbib]{mn2e}
\usepackage{graphicx}
\title[Mass-to-light Ratio of Ly$\alpha$ Emitters]{Mass-to-light
Ratio of Ly$\alpha$ Emitters: Implications of Ly$\alpha$ Surveys
at Redshifts $z=5.7$, $6.5$, $7$, and $8.8$}
\author[Elizabeth R. Fernandez and Eiichiro Komatsu]{Elizabeth R. Fernandez\thanks{beth@astro.as.utexas.edu} and Eiichiro Komatsu\\
Department of Astronomy, University of Texas at Austin,
1 University Station, C1400, Austin, TX 78712
}%
\begin{document}
\label{firstpage}
\maketitle
\begin{abstract}
 Using a simple method to interpret the luminosity function of
 Ly$\alpha$ 
 emitters, we explore properties of Ly$\alpha$ emitters from $5.7\le z\le
 8.8$ with various assumptions about metallicity and stellar
 mass spectra.  
 We constrain a mass-to-``observed light'' 
 ratio, $M_h/L_{band}$, where $M_h$ refers to the total mass  of the
 host halo, and $L_{band}$ refers to the observed luminosity of the
 source. For narrow-band surveys, $L_{band}$ is simply related to the
 intrinsic Ly$\alpha$ luminosity with a survival fraction of Ly$\alpha$
 photons, $\alpha_{esc}$. The  mass-to-``bolometric light'',
 $M_h/L_{bol}$, can also be deduced, once the metallicity and stellar
 mass spectrum are given.  The inferred
 $M_h/L_{bol}$ is more sensitive to metallicity than to the mass spectrum.
 We find the following constraints on a mass-to-light ratio of
 Ly$\alpha$ emitters from $5.7\le z \le 7$:
 $(M_h/L_{bol})(\alpha_{esc}\epsilon^{1/\gamma})^{-1}=21-38$, $14-26$, and $9-17$ for $Z=0$,
 $1/50$, and $1~Z_{\sun}$, respectively, where
 $\epsilon$ is the ``duty cycle'' of Ly$\alpha$ emitters, and
 $\gamma\sim 2$ is a local slope of the cumulative luminosity function,
 $N(>L)\propto L^{-\gamma}$, to which  the current data are sensitive.
 Only weak lower limits are
 obtained for $z=8.8$.
 Therefore, Ly$\alpha$ emitters
 are consistent with either starburst galaxies ($M_h/L_{bol}\sim 0.1-1$)
 with a smaller Ly$\alpha$ survival fraction,
 $\alpha_{esc}\epsilon^{1/\gamma}\sim 0.01-0.05$, or normal populations ($M_h/L_{bol}\sim 10$) if
 a good fraction of Ly$\alpha$ photons survived,
 $\alpha_{esc}\epsilon^{1/\gamma}\sim 
 0.5-1$. 
 We
 find no evidence for the end of reionization in the luminosity
 functions of Ly$\alpha$ emitters discovered in the current Ly$\alpha$
 surveys, including recent discovery of one Ly$\alpha$ emitter at $z=7$.
 The data are consistent with no
 evolution of intrinsic properties of Ly$\alpha$ emitters or neutral
 fraction in the intergalactic medium up to $z=7$.
 No detection of sources at $z=8.8$ does not yield a significant constraint yet.
 We also show that the lack of
 detection at $z=8.8$ does not rule out the high-$z$ galaxies being the
 origin of the excess near infrared background. 
 \end{abstract}
\section{Introduction}
\label{sec:introduction}
What was the universe like at high redshifts?  We are currently entering
a time where we can begin to observe this early time in the universe's
life by looking for galaxies at redshifts above 6.  

There are several indications that stars, and hence galaxies, existed at
this early time. We know that the universe was reionized early from
observations such as the polarized light of the cosmic microwave
background \citep{zaldarriaga:1997,kaplinghat/etal:2003,kogut/etal:2003,
page/etal:2006,spergel/etal:2006},
the Gunn-Peterson test towards quasars
\citep[e.g.][]{gunn/peterson:1965,fan/etal:2000,fan/etal:2001,fan/etal:2002,fan/etal:2004,becker/etal:2001,
oh/furlanetto:2005,goto:2006} and a gamma-ray burst
\citep{totani/etal:2006}, and the temperature of the intergalactic 
medium \citep{hui/haiman:2003}.    

In order to produce large scale reionization, an efficient and plentiful
source of ultraviolet photons was needed.  The first  
few generations of stars are very likely candidates, producing
ultraviolet photons efficiently.   
The Ly$\alpha$ forest shows that the universe was polluted with metals as
early as $z \sim 6$ \citep{songaila:2001, pettini/etal:2003,
ryanwebber:2006, simcoe:2006}, indicating even earlier star formation
and thus providing further support of early stars.  
In addition, a portion of the near infrared background could be the
redshifted light from the first stars, and may provide information
about them \citep{santos/bromm/kamionkowski:2002,
salvaterra/ferrara:2003, cooray/yoshida:2004, mii/totani:2005,madau/silk:2005,
fernandez/komatsu:2006,kashlinsky/etal:2005,kashlinsky/etal:2007}.   

It is therefore very likely that there is significant star formation
above $z>6$.  With the introduction of new, more powerful  
telescopes and deep field searches, an interesting question arises:  
do these first stars form galaxies that are bright enough to be seen
today?  Several deep field Ly$\alpha$ searches have been performed with
the ISAAC on the VLT \citep{willis/courbin:2005,willis/etal:2006,cuby/etal:2007}, the Mayall Telescope at Kitt Peak \citep{rhoads/etal:2004}, and the Subaru
telescope \citep{taniguchi/etal:2005,iye/etal:2006,kashikawa/etal:2006,shimasaku:2006}, probing the universe at redshifts of six and higher.   

In this paper, we present a simple method
to calculate the luminosity function of high-$z$ galaxies, and
compare this with the results of Ly$\alpha$ searches to
constrain properties of Ly$\alpha$ emitters.
In \S~\ref{sec:Formalism}, we explain our
method to calculate the luminosity function of high-$z$ galaxies 
with a single free parameter, a mass-to-observed light ratio, $M_h/L_{band}$.
 In \S~\ref{sec:Observations}, we review the current generation high redshift 
galaxy surveys and their data.  In \S~\ref{sec:interpretation} we
model stellar populations to obtain actual physical quantities about the
galaxies that we observe.   We compare our results to previous work in
\S~\ref{sec:compare} and conclude in \S~\ref{sec:conclusions}.  We use
the latest WMAP 3 parameters of $\Omega_b=0.0422$, $\Omega_m =0.241$,
$\Omega_{\Lambda}=0.759$, $\sigma_8=0.761$, $h=0.732$, and $n_s=0.958$
\citep{spergel/etal:2006}. 

\section{A Simple Model of Galaxy Counts}
\label{sec:Formalism}
\subsection{Justification for a simplified approach}
\label{sec:justification}
The simplest way to predict the cumulative luminosity function of 
galaxies is to count the
number of haloes available in the universe above a certain mass,
\begin{equation}
N(>L) = V(z)\int_{M_h(L)}^\infty\frac{dn}{dM_h}dM_h,
\label{eq:simplemodel}
\end{equation}
where $V(z)$ is the survey volume, $dn/dM_h$ the comoving number density
of haloes per unit mass range, and $M_h$ the total mass of a halo.

The cumulative number density of haloes, $\int_{M_h}^\infty 
dM_h~dn/dM_h$, is shown
in the bottom panel of Figure~\ref{fig:STdndm}. 
In order to calculate the cumulative luminosity function, one may 
simply ``stretch'' the horizontal axis of this figure by
dividing $M_h$ by a suitable factor that converts the mass to
luminosity: a mass-to-light ratio, $M_h/L$.

This model is admittedly oversimplified, and is indeed simpler than what
is already available in the literature. For example, one can stretch not only
the horizontal axis (i.e., mass), but also the vertical axis of the 
cumulative mass function as $\int dM_h~dn/dM_h\rightarrow
\epsilon\int dM_h~dn/dM_h$, 
where $\epsilon$ is often called a ``duty cycle''
\citep[e.g.,][]{haiman/spaans/quataert:2000}.

The vertical stretch would be required when the average lifetime of Ly$\alpha$
emitting galaxies, $\tau_g$, 
is shorter than the age of the universe, in which case the
number count should be given by the time derivative of the mass function,
$\int dM_h~dn/dM_h\rightarrow 
\int dM_h~\tau_g(M_h)~d^2n/(dM_hdt)$. The vertical stretch
parameter is thus
given approximately by $\epsilon\approx \tau_g/t_{\rm univ}$.

Since the statistical power of the current data is not yet strong enough to
constrain both the horizontal and vertical stretch parameters
simultaneously, these 
parameters 
are strongly degenerate
\citep{dijkstra/etal:2006b,stark/loeb/ellis:2007}. 
They are completely degenerate when $N(>L)$
follows a single power law, $N(>L)\propto L^{-\gamma}$.
The degeneracy line is given by
$(M_h/L)\epsilon^{-1/\gamma}=\mbox{constant}$; thus,
the inferred $M_h/L$ and $\epsilon$ are positively
correlated: 
the smaller the $\epsilon$ is, the smaller the inferred $M_h/L$ becomes.

In order to lift this degeneracy, therefore,
it is essential to detect the deviation of $N(>L)$
from a power law. Since the cumulative mass
function in $5.7\le z\le 8.8$ begins to decline exponentially with mass at 
a few times $10^{11}~M_{\sun}$ at $z\sim 6$ to $10^{10}~M_{\sun}$ at $z\sim 9$
(see Fig.~\ref{fig:STdndm}), an accurate
determination of the bright end of the luminosity function at 
$L_{Ly\alpha}\ga 10^{42}~{\rm erg~s^{-1}}\times (100~L_{Ly\alpha}/M_h)$ 
at $z\sim 9$ to 
$10^{43}~{\rm erg~s^{-1}}\times (100~L_{Ly\alpha}/M_h)$ at
$z\sim 6$ would be required to lift the degeneracy.
(Note that $M_h/L$ is always quoted in units of $M_{\sun}L_{\sun}^{-1}$,
where
$L_{\sun}=3.8\times 10^{33}~{\rm erg~s^{-1}}$.)
The bright end of the luminosity function is not constrained very well by the
existing surveys (see Fig.~\ref{fig:numberFields}). One would need a
larger survey area for a better determination of the bright end of the
luminosity function.  (See section \ref{sec:lband} for further discussion on constraining the bright end of the luminosity function.)

We have chosen to work with
$\epsilon=1$, which is allowed by the existing data
\citep{dijkstra/etal:2006b,stark/loeb/ellis:2007}.
As we show in this paper, this assumption does provide
reasonable and useful results. 
We then use the degeneracy line,
$(M_h/L)\epsilon^{-1/\gamma}=\mbox{constant}$, to incorporate the effect
of $\epsilon$ into the inferred constraints 
on properties of Ly$\alpha$ emitters.

A further improvement to the model can be made by taking into account
the fact that a relation between the luminosity and halo mass is not
unique, but has some dispersion. One may include this by using a
conditional probability of luminosity given the halo mass, $P(L|M_h)$, as 
$N(>L) = V(z)\int dM_h~P(L|M_h)~dn/dM_h$ with $P(L|M_h)$ given by, e.g.,
a log-normal distribution \citep{cooray/milosavljevic:2005}.
Once again, the current data cannot constrain the extra parameters
characterizing $P(L|M_h)$, except for its first-order moment, a 
mass-to-luminosity
ratio. In our simplified approach we take it to be a delta
function, $P(L|M_h)=\delta^D[M_h-(M_h/L)L]$, which gives
equation~(\ref{eq:simplemodel}). 

In this paper we shall use the Sheth-Tormen formula 
for $dn/dM_h$
\citep{sheth/mo/tormen:2001,sheth/tormen:2002}.  
Since the Press-Schechter mass function 
\citep{press/schechter:1974} tends to underestimate the number of haloes
in the high mass range, the mass-to-light ratio inferred from the
Press-Schechter mass function would be smaller than that from the
Sheth-Tormen formula. We have found that the mass-to-light inferred from the 
Press-Schechter mass function is smaller by a factor of two.

The volume, $V(z)$, is found by multiplying
the comoving volume element, $d^2V/dzd\Omega$, by the depth of the
survey (found by integrating over redshift) and the survey area on the
sky. 
 For narrow-band surveys,
the redshift integral can be approximated as $\Delta z$.   
 The comoving volume element is given by
\begin{equation}
\frac{d^2V}{dzd\Omega} =
 \frac{cd_L^2(z)}{1+z}\left(-\frac{1}{H(z)(1+z)}\right) ,  
\end{equation}
where $d_L$ is the proper luminosity distance.  

\subsection{Basic formalism}
From equation~(\ref{eq:simplemodel}) we can derive
the number of galaxies observed above a certain flux density as:   
\begin{eqnarray}
\label{eq:Fdndm}
\nonumber
& &\int^{\infty}_{F_{limit}} \frac{d^2N}{dFd\Omega}dF \\
\nonumber
&=& \int dz
\frac{d^2V}{dzd\Omega}\int^{\infty}_{F_{limit}}\frac{dn}{dM_h}\frac{dM_h}{dF}
\vartheta (M_h-M_{min}(z))dF\\  
&\approx &
\Delta z \frac{d^2V}{dzd\Omega}\int^{\infty}_{F_{limit}}\frac{dn}{dM_h}
\frac{dM_h}{dF}\vartheta (M_h-M_{min}(z)) dF. 
\end{eqnarray}
where $M_h$ is the halo mass. 

\begin{figure}
\centering \noindent
\includegraphics[width=8cm]{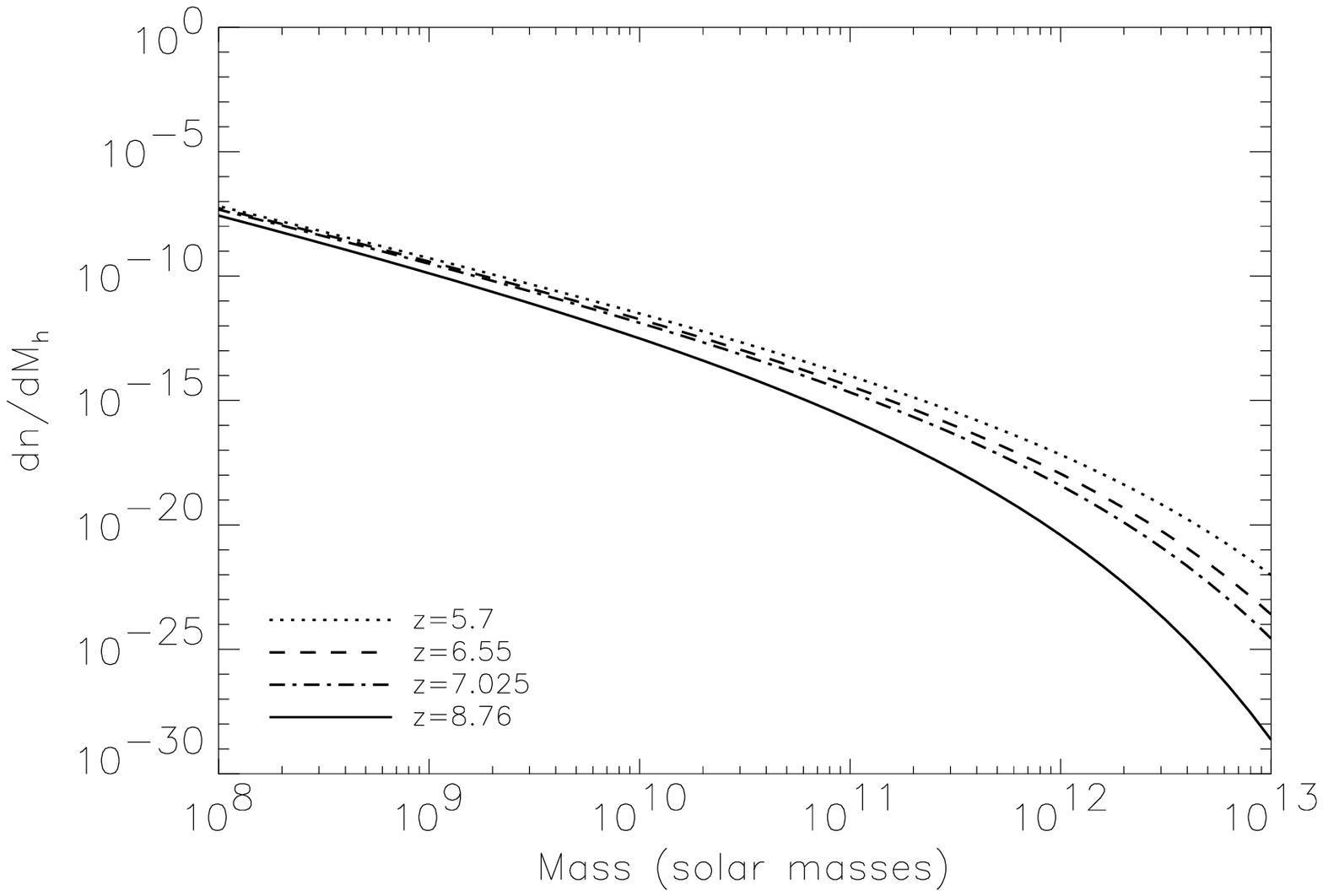}
\hspace{5mm}
\includegraphics[width=8cm]{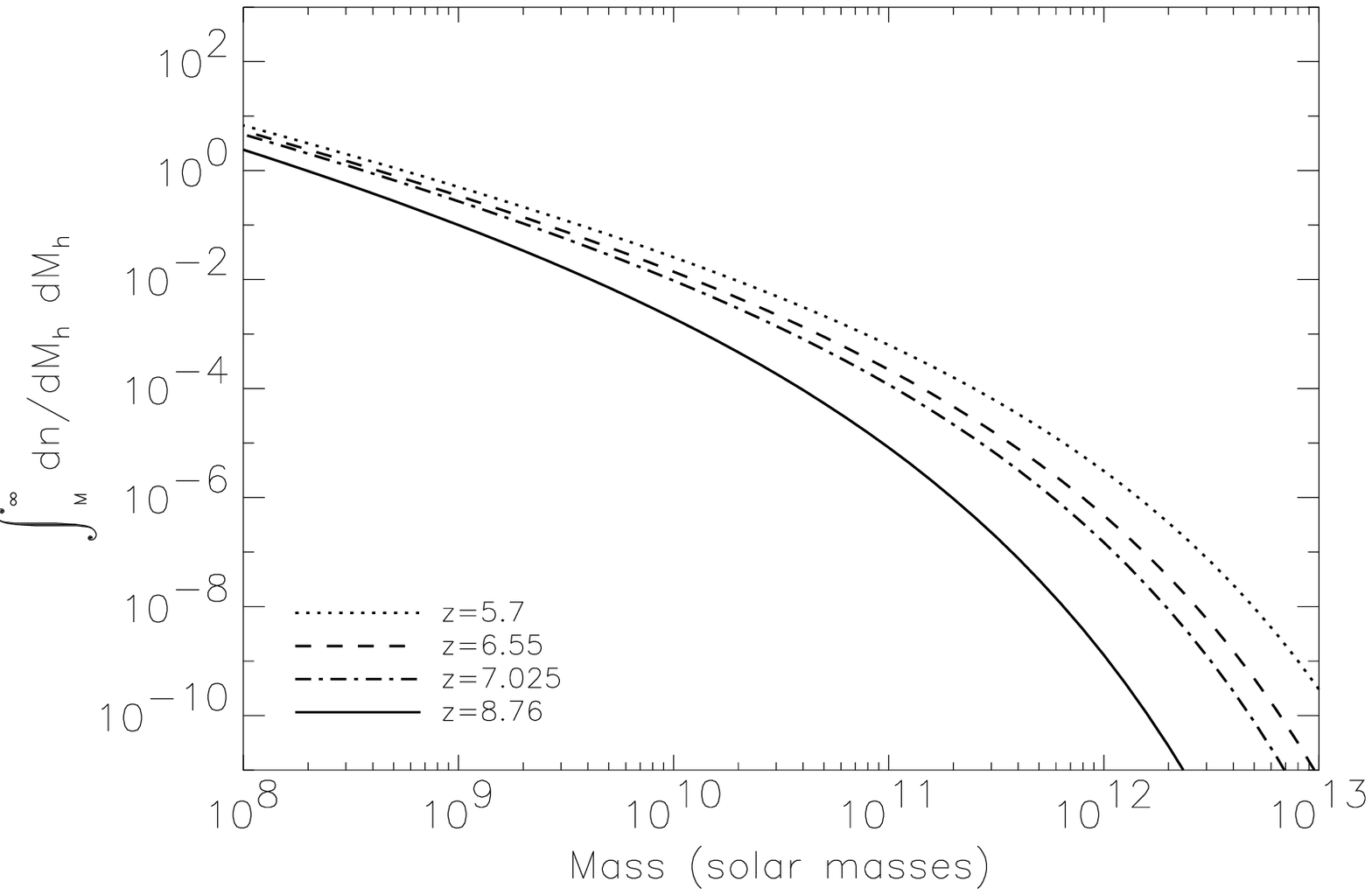}
\caption{%
 Mass function.
 The dotted, dashed, dot-dashed and solid lines show $z=5.7$, 6.55,
 7.025, and 8.76, respectively. These are the redshifts of narrow-band
 surveys that we consider in this paper (see Table~\ref{table:surveyprop}).
 ({\it Upper Panel}) The Sheth-Tormen mass function as a function of
 halo mass, $dn/dM_h$, in units of comoving ${\rm Mpc}^{-3}~{\rm M}_{\sun}^{-1}$ ({\it Lower Panel}) The cumulative mass function,
$\int_{M_h}^\infty dM_h~dn/dM_h$, in units of comoving ${\rm Mpc}^{-3}$.
 The cumulative luminosity function can be calculated 
 by stretching the horizontal axis of this figure by
 dividing $M_h$ by a suitable factor that converts the mass to
 luminosity: a mass-to-light ratio, $M_h/L$.
}
\label{fig:STdndm}
\end{figure}

Not all dark matter haloes will be forming
stars - only haloes with a mass above some critical minimum mass
($M_{min}$).  This is represented by the function
$\vartheta(M_h-M_{min}(z))$, which is zero if the halo mass is smaller
than $M_{min}$ and unity if it is larger than or equal to $M_{min}$. The
minimum mass is only theoretically known, and we use the virial mass of
a $10,000$ K halo,  
$M_{min} = 0.94\times 10^8~M_\odot[(1+z)/10]^{-3/2}$.
However, given the current sensitivity of telescopes, it is unlikely
that a halo of mass $M_{min}$ will be bright enough to be 
seen, unless the mass to light
ratio of galaxies is unusually small.  Therefore, $M_{min}$ is
irrelevant to our analysis presented in this paper and our conclusion is
independent of the actual value of $M_{min}$.

In deriving equation~(\ref{eq:Fdndm}),
we have made an assumption that each dark matter halo above $M_{min}$
hosts one galaxy.  
This is a valid assumption at high redshifts, as massive haloes such as
groups ($M \ga 10^{13} M_{\odot}$)
and clusters  ($M \ga 10^{14} M_{\odot}$) of galaxies 
hosting multiple galaxies are extremely rare (see
Fig. \ref{fig:STdndm}).  If we were to assume a field size equal to the largest survey area discussed in this paper (that of the LALA survey, 1296 arcmin$^2$), and the widest redshift range (that of the Subaru survey at z$\sim 7$), the number of haloes above a mass of $10^{13}M_{\odot}$ can be found by using equation \ref{eq:simplemodel}.  At a redshift of 1, there would be 32 haloes larger than this mass in the field.  At higher redshifts ($z=3$,$5$, and $7$), there would be less than one such massive halo in the field (0.98, 1$\times10^{-3}$, and $7.7\times10^{-8}$, respectively.)  Thus at high redshifts, it is safe to assume that there are no groups or clusters observed.  (Larger surveys \citep{ouchi/etal:2005} have observed protoclusters).
This property makes it possible to model the
luminosity function of high-$z$ galaxies without complications arising
from galaxy formation processes.  Some observations (i.e. \citet{ouchi/etal:2004}) show that some dark matter haloes host more than one Ly$\alpha$ emitter.  However, since galaxy occupation number at high redshifts is not well known, we will just assume one galaxy per halo.

The most important uncertainty in our model is that
not all galaxies are seen as Ly$\alpha$
emitters.
Some galaxies do not produce as many Ly$\alpha$ photons as the others do
because of dust extinction in galaxies themselves 
and scattering in the IGM. 
Therefore we could have assumed that there is less than one Ly$\alpha$ emitter
per halo; however, this effect can be modeled effectively by
introducing a Ly$\alpha$ survival fraction, $\alpha_{esc}$.   
This parameter quantifies the fraction of Ly$\alpha$ photons that escaped
from a halo {\it and} the IGM.  Therefore, as $\alpha_{esc}$ increases, a galaxy is seen as more luminous intrinsically.

In summary, we model the galaxy number counts by placing one galaxy per
halo with only a fraction of photons escaping from galaxies and the IGM.

\subsection{Mass-to-``observed light'' ratio}
Our formulation is now reliant on how we relate the flux density of a galaxy to
its mass.  

The flux density of galaxies depends on two things: the luminosity distance to
galaxies, and a mass-to-``observed light'' ratio, which will relate the total
mass of the halo (including dark matter) to the luminosity that is
actually observed.  The flux density of a galaxy observed by a certain 
instrument
is found by: 
\begin{equation} 
\label{eq:fluxlum}
F=\frac{L_{band}/M_h}{4\pi d_L^2(z)\Delta \nu_{obs}} 
M_h ,
\end{equation}
where $\Delta \nu_{obs}$ is the bandwidth of the instrument (which we have assumed to have an ideal rectangular bandpass) and 
\begin{equation}
L_{band}=\int^{\nu_{2,obs}(1+z)}_{\nu_{1,obs}(1+z)}d \nu L_{\nu} ,
\end{equation}
is the observed luminosity within a certain bandwidth of the instrument,
 $L_\nu$ the rest-frame luminosity per unit rest-frame frequency, 
and $\nu_2$ and $\nu_1$ the frequency
 limits of the survey. 

We assume that the mass to light ratio is independent of mass, 
$dM_h/dF = M_h/F$. This approximation is well justified, as the current
surveys are probing a limited mass range. Total and stellar masses are still unknown for Ly$\alpha$ emitters.  Studies of the stellar masses using SED fitting have just begun \citep{mobasher/etal:2005,nilsson/etal:2007}.  In the future when the
observations of Ly$\alpha$ emitters can cover a wide mass range, one may use a
parametrized model, e.g., $L\propto M^\beta$, to improve fits.
For the present purpose additional parameters are unnecessary.  

Using equation (\ref{eq:fluxlum}),
equation (\ref{eq:Fdndm}) can be rewritten with respect to 
$M_h/L_{band}$:   
\begin{eqnarray}
\label{eq:ML}
\nonumber
& &\int^{\infty}_{F_{limit}} \frac{d^2N}{dFd\Omega}dF \\
\nonumber
&=& 4\pi d_L^2(z)  \frac{dV}{dzd\Omega}
\Delta z  \Delta \nu_{obs}
\frac{M_h}{L_{band}}
\int^{\infty}_{F_{limit}}\frac{dn(M_h(F))}{dM_h} dF\\
& &\times\vartheta(M_h-M_{min}(z)).
\end{eqnarray}
To evaluate $dn/dM$ for a given $F$, we use equation~(\ref{eq:fluxlum})
to convert $F$ to $M_h$. 
Once again, 
$M_h/L_{band}$ is independent of $M_h$, and 
$M_h$ is almost always greater than $M_{min}$, 
and thus almost always 
$\vartheta(M_h-M_{min}(z)) = 1$.  The only unknown quantity in this
equation is $M_h/L_{band}$, while $\Delta \nu_{obs}$, $F_{limit}$, and
$\Delta z$ are given by the survey properties.  
In other words, 
$M_h/L_{band}$
is the parameter that should be measured from the observational data directly.

\section{Narrow-band Ly$\alpha$ surveys: Observations}
\label{sec:Observations}
\begin{table*}
\begin{minipage}{16cm}
\caption{%
Survey parameters taken from
\citet{shimasaku:2006}$^a$, 
\citet{rhoads/etal:2004}$^b$,
\citet{taniguchi/etal:2005}$^c$,
\citet{kashikawa/etal:2006}$^d$,
\citet{iye/etal:2006}$^e$,
\citet{willis/courbin:2005}$^f$,
\citet{willis/etal:2006}$^g$, and
\citet{cuby/etal:2007}$^h$.
}%
\begin{center}
\begin{tabular}{|l|l|l|l|l|l|l|l|}
\hline
Name of survey & Telescope & Central $\lambda$ (\AA) & Bandwidth
 (\AA)& Central $z$ & Redshift Range& Area (arcmin$^2$) & Ref\\
\hline
Subaru Deep Field & Subaru & 8150 & 120 & 5.7 & $5.64-5.76$& 725&a \\
LALA & Mayall & 9182 & 84 & 6.55 & $6.516-6.586$ & 1296&b\\
Subaru Deep Field  &Subaru & 9196 & 132 & 6.56 & $6.508-6.617$ &
			 876&c, d\\
Subaru Deep Field  &Subaru & 9755 & 200 & 7.025 & $6.94-7.11$ & 876&e\\
ZEN & VLT & 11900  & 89.5 & 8.76 & $8.725-8.798$ & 4 & f, g\\
ISAAC ext & VLT & 11900 & 89.5 & 8.76 & $8.725-8.798$ & 31&h\\
\hline
\end{tabular}
\label{table:surveyprop}
\end{center}
\end{minipage}
\end{table*}

Several telescopes are now powerful enough to attempt to locate galaxies
at $z\ga 6$.  One effective method of locating high redshift
galaxies is to use narrow-band filters to detect Ly$\alpha$ emission
from a small range of redshifts.  Ly$\alpha$ emitters tend to have 
large equivalent widths, with their line intensities
significantly higher than the continuum emission.  In order to test that the
galaxy is indeed a 
high redshift galaxy, it must be undetected at optical
wavelengths, and follow-up spectroscopy may be employed if possible.  At
times, when no continuum is visible, the asymmetric profile of the
Ly$\alpha$ line can be used to identify a Ly$\alpha$ emitter.  

A large number of such narrow-band 
Ly$\alpha$ searches have been carried out on blank fields at $z=5.7$
\citep{rhoads/malhotra:2001,rhoads/etal:2003,maier/etal:2003,hu/etal:2004,ouchi/etal:2005,ajiki/etal:2003,ajiki/etal:2004,ajiki/etal:2006,shimasaku:2006,murayama/etal:2007},   
$z=6.5$
\citep{kodaira/etal:2003,taniguchi/etal:2005,kashikawa/etal:2006}, $z=7$
\citep{iye/etal:2006}, and $z=8.8$
\citep{willis/courbin:2005,willis/etal:2006,cuby/etal:2007}.
\citet{kurk/etal:2004} used an alternative technique, a slitless-grism
spectroscopy survey, which has also yielded a successful detection of 
a Ly$\alpha$ emitter
at $z=6.5$. 
\citet{martin/sawicki:2004} performed a multi-slit windows search
at $z=5.7$.

In this paper we use the observational data from 
six narrow band Ly$\alpha$ searches: two from
the Infrared Spectrometer and Array Camera (ISAAC) on the Very Large
Telescope (VLT), three from Subaru, and one from the Mayall Telescope at
Kitt Peak. The basic parameters of these surveys are summarized in Table~\ref{table:surveyprop}.

 Two surveys at $z=5.7$ and $z=6.56$ were taken at the Subaru telescope
 using the Suprime-Cam.  The surveys covered 725 and 875 arcmin$^2$,
 respectively.  These surveys had a limiting flux density of $145$ nJ.  Candidates were found down to a limiting magnitude of $26.0$ (at a detection limit of $3\sigma$ for z=5.7 and $5\sigma$ for z=6.56)  using a $2''$ aperture.  At
 $z=5.7$, there were 89 Ly$\alpha$ emitter candidates, 63 were followed
 up by spectroscopy, and 34 Ly$\alpha$ emitters 
have been confirmed
 \citep{shimasaku:2006}.  At $z=6.56$, 
there were 58 Ly$\alpha$ emitter candidates,  
53 were followed up by spectroscopy, and
17 Ly$\alpha$ emitters 
have been confirmed
 \citep{taniguchi/etal:2005,
 kashikawa/etal:2006}.  From these detections, they were able to fit a
 Schechter function, $\phi(L)dL =
 \phi^*(L/L^*)^{\alpha}\exp(-L/L_*)dL/L^*$, with 
 parameters given in Table \ref{table:lumfunc}.  

The Large Area Lyman Alpha (LALA) survey searched for galaxies at a
redshift of around 6.55 \citep{rhoads/etal:2004}.  This survey was
conducted on the 4-m Mayall Telescope at Kitt Peak.  A narrow-band 1296
arcmin$^2$ image was taken of the Bo\"otes field.  The limiting flux density
was 700~nJ.  Three candidates were located down to a limiting magnitude of $24.3$ ($5\sigma$) with a $1''.02$ aperture -- one of which was confirmed
spectroscopically as a Ly$\alpha$ emitter. 

Another survey using the Subaru telescope
covered an area of 876 arcmin$^2$ at $z=7.025$.  They used the narrow
band filter NB973 on the Subaru Suprime-Cam and had a limiting flux density of
398~nJ. 
There were 2 Ly$\alpha$ emitter candidates down to a limiting magnitude of $24.9$ ($5\sigma$) with a $2''$ aperture  -- one of which
was confirmed
spectroscopically as a Ly$\alpha$ emitter at $z=6.96$
\citep{iye/etal:2006}. 

ZEN, which stands for $z$ equals nine, is a narrow J-band mission using
the ISAAC on the VLT.  Its central redshift is 8.76 and it covers a
field size of 4 arcmin$^2$ 
down to a limiting flux density of 302~nJ.  They
searched for galaxies in the Hubble Deep Field South that displayed an
excess in the narrow band in comparison to the J-band ($J_s-NB \geq
0.3$) and that were undetected in the optical.  No galaxies were found down to a limiting magnitude of $25.2$ ($5\sigma$) with a $0''.7$ aperture
\citep{willis/courbin:2005, willis/etal:2006}. 

\citet{cuby/etal:2007} did a followup narrow-band search, using the
ISAAC at the VLT, with a larger field of view (hereafter referred to as
the ISAAC ext).  They imaged seven fields in the Chandra Deep Field South
that totaled 31 arcmin$^2$ down to a flux density of 1740~nJ.  They also
detected no galaxies within the fields down to a limiting magnitude of $23.3$ ($5\sigma$) with a $1''$ aperture. 

\section{Properties of Ly$\alpha$ emitters}
\label{sec:interpretation}
\subsection{Extracting a mass-to-``observed light'' ratio}
\label{sec:lband}
\begin{table*}
\begin{minipage}{10cm}
\caption{%
 Best-fitting Schechter parameters from
 \citet{kashikawa/etal:2006}. These data are plotted in
 Figure~\ref{fig:numberFields}.
}%
\begin{center}
\begin{tabular}{|l|l|l|l|}
\hline
Redshift  & $\alpha $  & $\log_{10}(L^*/h^{-2}_{70}$ erg s$^{-1}) $ & $ \log_{10}(\phi^*/h^3_{70}~$Mpc$^{-3})$  \\
\hline
6.5  & $-2.0$ & 42.74 & $-3.14$\\
  &  $-1.5$ & 42.60 & $-2.88$\\
  & $-1.0$ & 42.48 &$-2.74$\\
5.7 & $-2.0$ & 43.30 & $-3.96$\\
  &  $-1.5$ & 43.04&$-3.44$\\
  & $-1.0$ & 42.84&$-3.14$\\
\hline
\end{tabular}
\label{table:lumfunc}
\end{center}
\end{minipage}
\end{table*}


\begin{figure*}
\centering \noindent
\includegraphics[width=8cm]{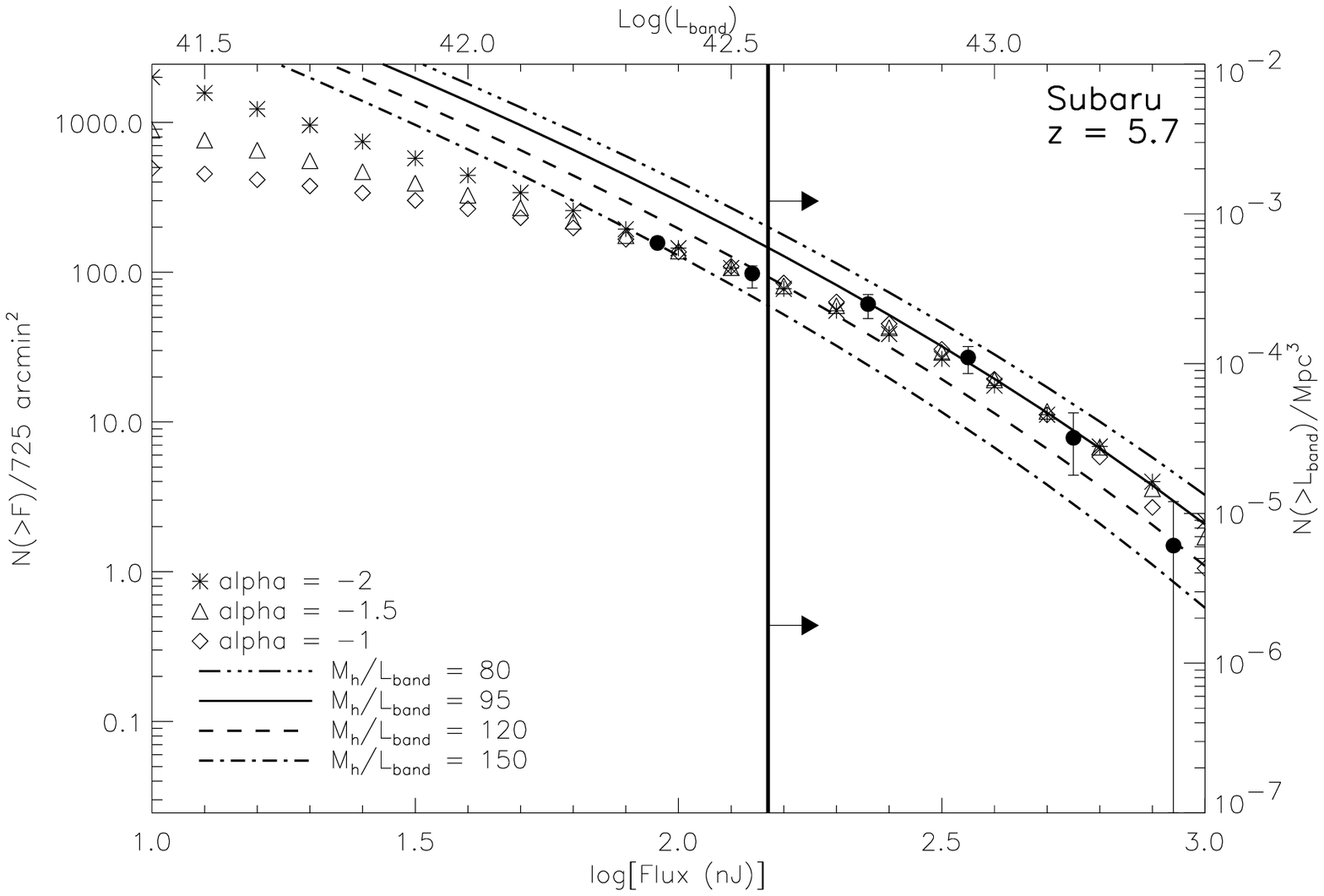}
\hspace{5mm}
\includegraphics[width=8cm]{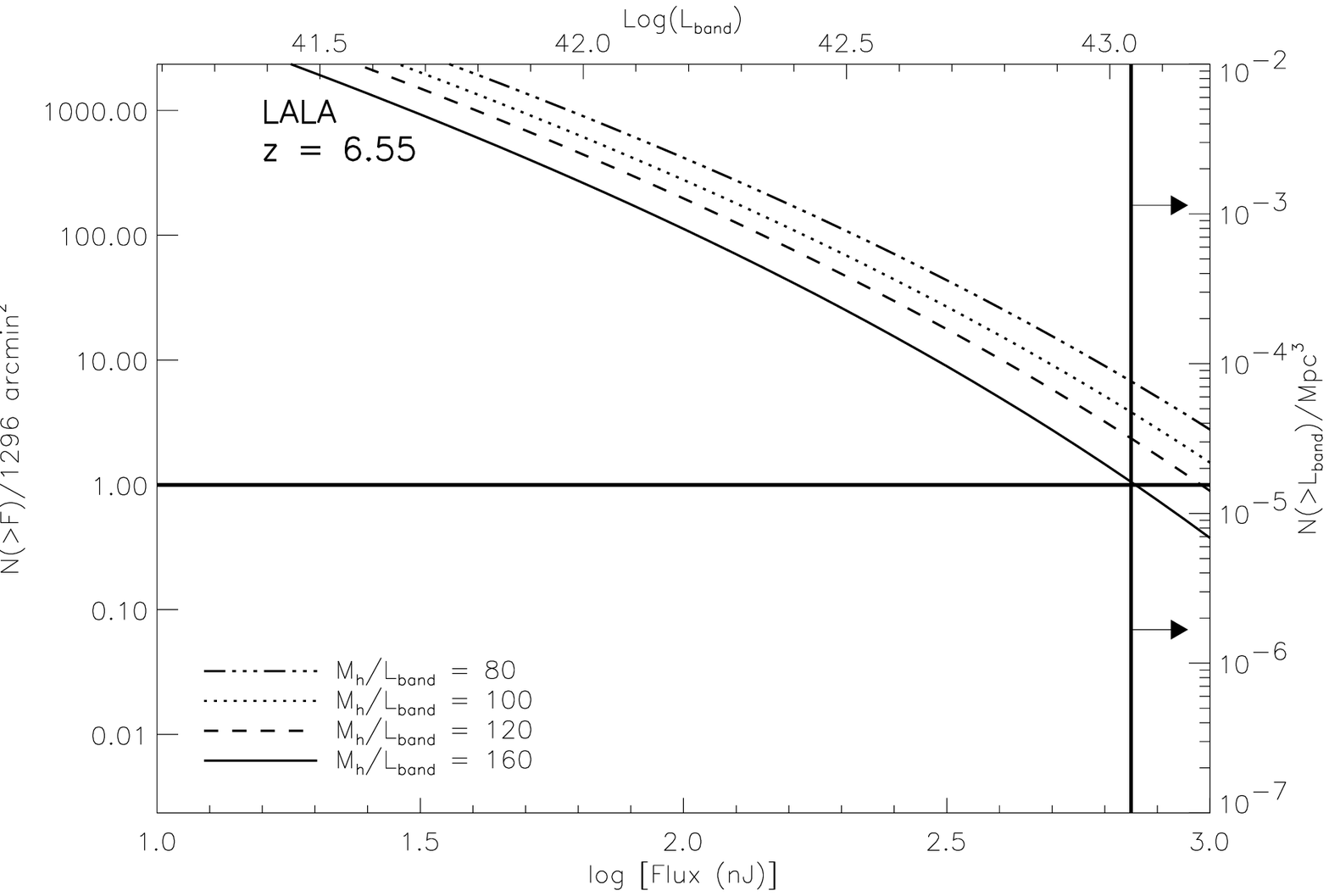}\\
\vspace{5mm}
\includegraphics[width=8cm]{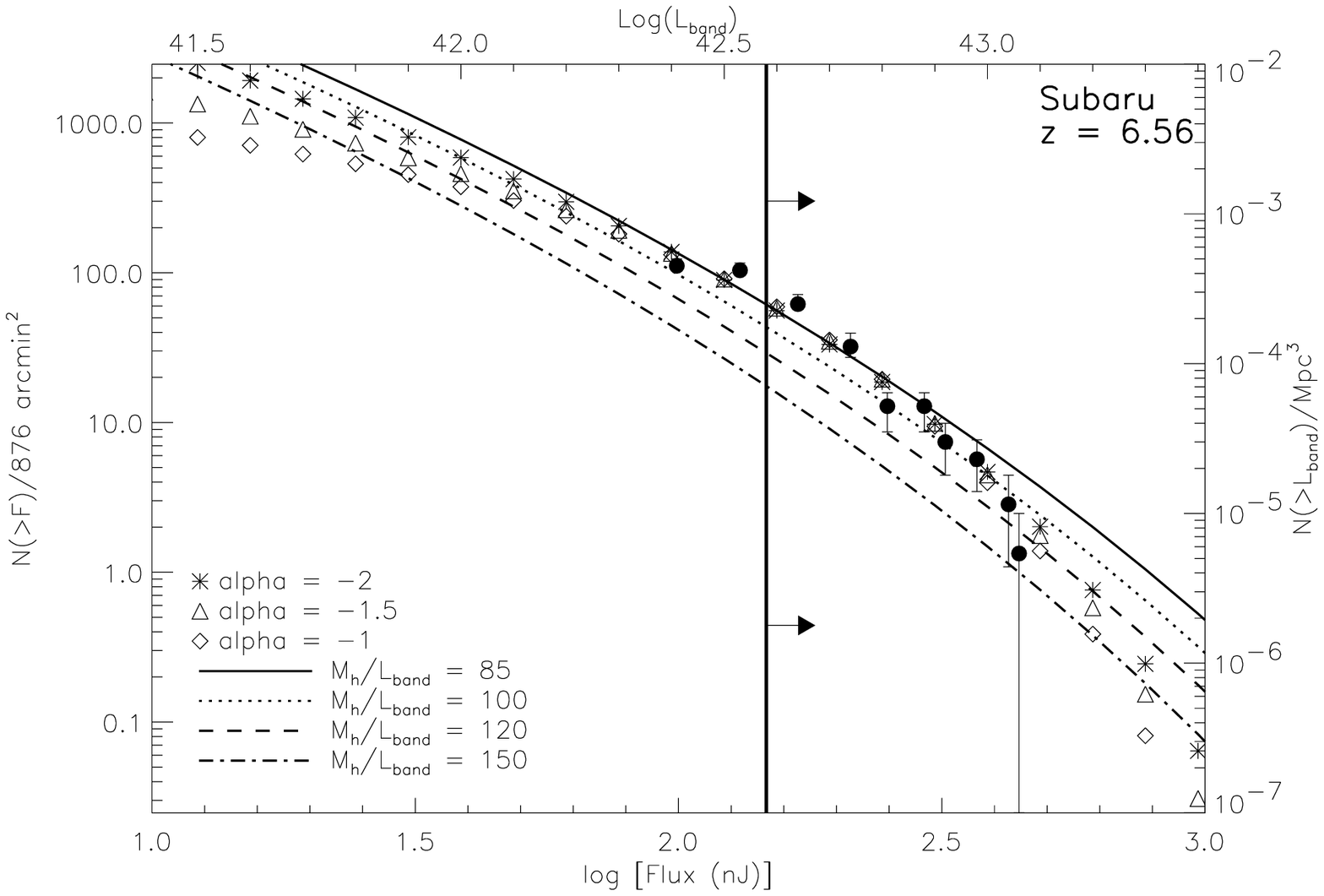}
\hspace{5mm}
\includegraphics[width=8cm]{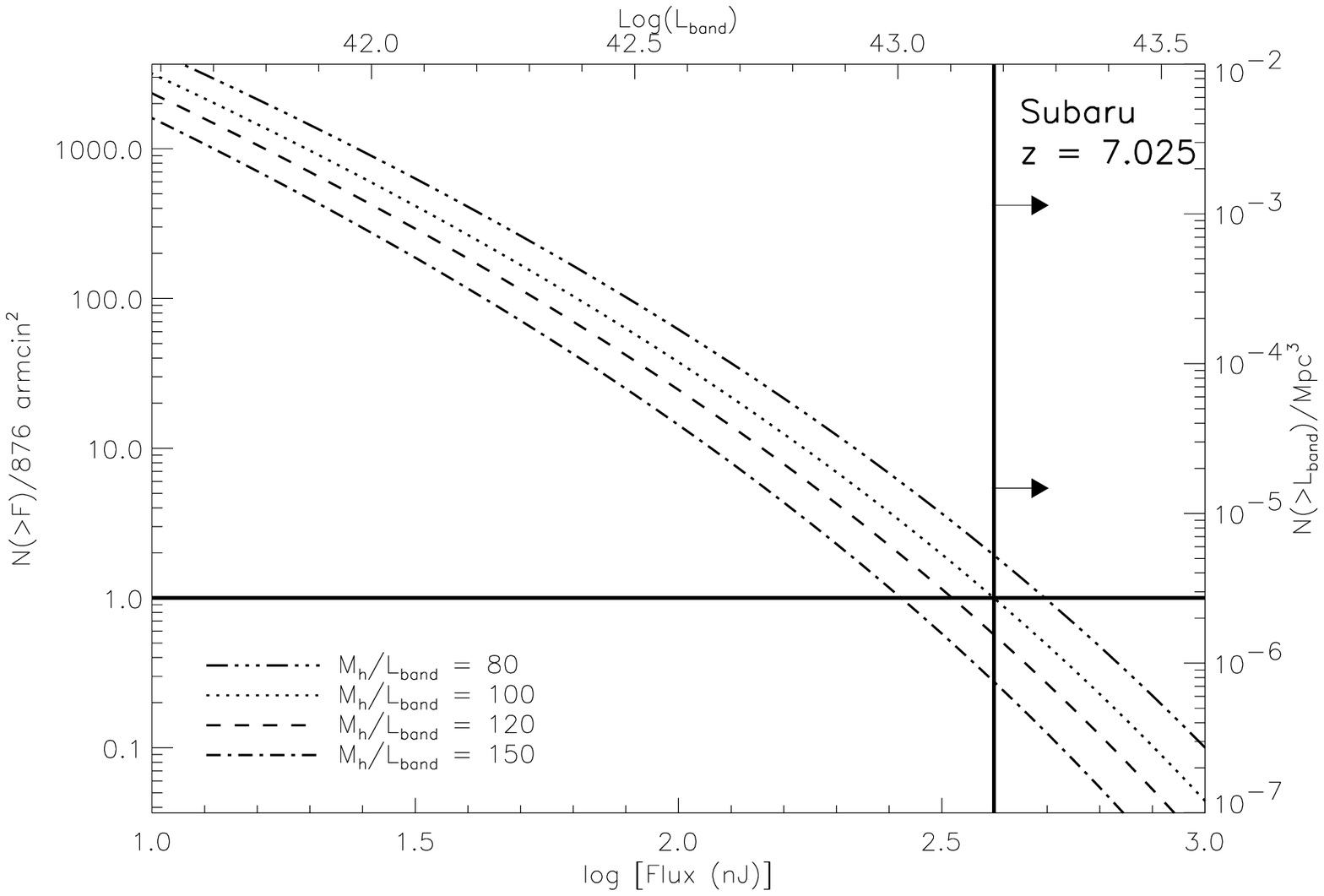}\\
\vspace{5mm}
\includegraphics[width=8cm]{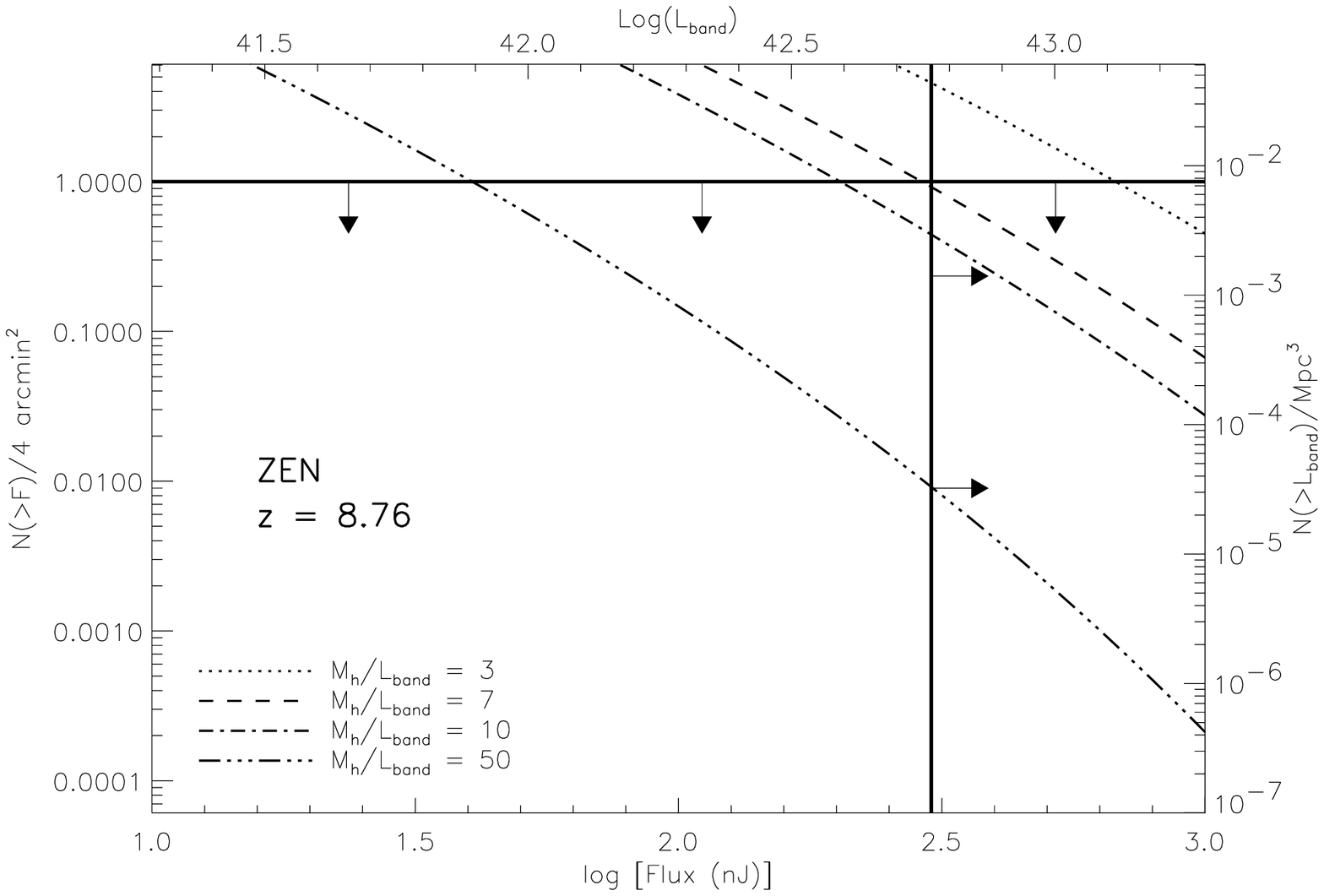}
\hspace{5mm}
\includegraphics[width=8cm]{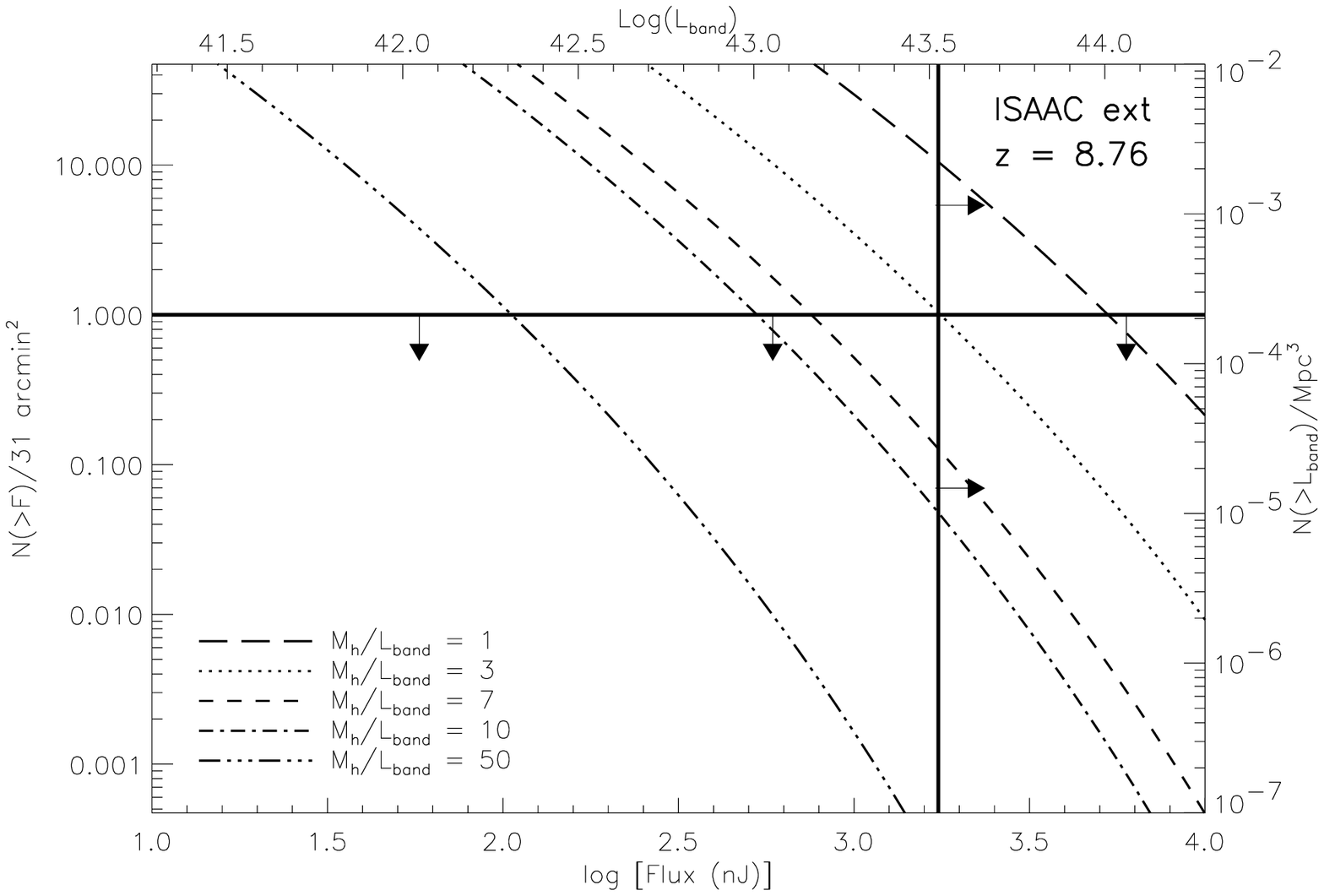}
\caption{%
 The observed luminosity function of Ly$\alpha$ emitters constrains their
 mass-to-``observed light'' ratio. Each panel shows the cumulative
 number of sources detected in each field above a certain flux density, 
 $N(>F)$. (The flux density limits of each survey are indicated by the vertical lines
 with right arrows.) The bottom and top axis show the measured flux density and
 luminosity (in erg s$^{-1}$), respectively, while the left and right show the
 number of sources per field and per comoving Mpc$^{3}$, respectively.
 The mass-to-``observed light'' ratio, $M_h/L_{band}$, is fit to
 each of the luminosity functions derived from various narrow-band
 surveys. Here, $L_{band}$ refers to the light that falls within the band of 
 instruments, which is mostly the Ly$\alpha$ line.
 Curves in each panel show predictions with various $M_h/L_{band}$.
 ({\it Upper Left}) The Subaru field at $z=5.7$ \citep{shimasaku:2006}.
 The stars, triangles and diamonds show their Schechter-fit to the
 luminosity function with $\alpha=-2$, $-1.5$ and $-1$,
 respectively. The scatter between symbols shows
 uncertainty, and they diverge mostly below the flux density limit, as expected.  The solid circles are the data, with the error bars showing Poisson error. 
 ({\it Upper Right}) The LALA field at $z=6.55$
 \citep{rhoads/etal:2004}. One Ly$\alpha$ emitter was found, and the horizontal line
 shows $N(>F)=1$ per field.
 ({\it Middle Left}) The Subaru field at $z=6.56$
 \citep{taniguchi/etal:2005,kashikawa/etal:2006}. 
The meaning of symbols is the same as in
 the upper left panel.  Data is from both the photometric and spectroscopic samples.  The circles show data corrected for detection completeness.
 ({\it Middle Right}) The Subaru field at
 $z=7.025$ \citep{iye/etal:2006}. One Ly$\alpha$ emitter was found.
 ({\it Bottom Left}) The ZEN field at $z=8.76$
 \citep{willis/courbin:2005,willis/etal:2006}. No sources were found, and the horizontal line
 with lower arrows shows $N(>F)<1$ per field.
 ({\it Bottom Right}) The ISAAC ext field at $z=8.76$
 \citep{cuby/etal:2007}. No sources were found.
}
\label{fig:numberFields}
\end{figure*}

Can we infer anything about properties of the Ly$\alpha$ emitters
discovered in these narrow-band surveys?
Is the lack of detections in the fields above $z>7$ expected, or should
we expect many more galaxies?   

In equation (\ref{eq:ML}), the only free parameter was
the mass-to-``observed light'' ratio, 
$M_h/L_{band}$.  Therefore, we vary $M_h/L_{band}$ to give a model
that is consistent with observations.  
The number of galaxies drops as
$M_h/L_{band}$ increases.  (See Fig. \ref{fig:numberFields}.)
As the mass-to-light ratio increases, the star formation is spread out
over a longer period of time.
Therefore, the galaxies are dimmer because less
stars are shining at any given time.   

We find that the Subaru data at $z=5.7$ and $z=6.56$ are 
consistent with no evolution 
of properties of Ly$\alpha$ emitters or the IGM
opacity.
The evolution in the number
density of Ly$\alpha$ emitters can be explained solely
by the evolution of the halo mass function. These points have been made already by 
\citet{malhotra/rhoads:2004,dijkstra/etal:2006a,mcquinn/etal:2006}. 
At $z=5.7$
we find that $M_h/L_{band}=95-120$
fits the Subaru data from \citet{shimasaku:2006}, with lower values favored near the flux density limit, where it is harder to correct for sample completeness.
At $z=6.56$ a slightly (20\%) smaller value, $M_h/L_{band}=85-100$,
fits the Subaru data from \citet{kashikawa/etal:2006}.
(These are fit for the values of $L_{band}$ below $10^{43}$ erg s$^{-1}$
to the flux density limit, for most of the observational data fall between these
limits.) 

The high value of $M_h/L_{band}=95-120$ at $z=5.7$ seems to fall outside the Poisson error at $z=6.56$ near the flux density limit, although at higher fluxes its statistical significance is more questionable.
If we take this $\sim 20$\% decrease in $M_h/L_{band}$ seriously, an
interesting conclusion may be drawn. First of all, 
the {\it decrease} in $M_h/L_{band}$ from $z=5.7$ to $z=6.56$
is qualitatively inconsistent with the evolution of neutral fraction 
in the IGM.
If the IGM was more neutral (i.e., less ionized) in the past, 
we should observe the {\it
increase} in  $M_h/L_{band}$ at higher $z$. 
The evolution in $M_h/L_{band}$ may be even more significant than it
looks now, once the redshift effect is taken into account. 
Since the survey at $z=6.56$ collects less
photons than that at $z=5.7$ for a given bandwidth of the instrument,
one must take into account the bandwidth properly before making 
a quantitative comparison between $M_h/L_{band}$ from two different redshifts.
 We shall perform this analysis more
carefully in section~\ref{sec:anomaly}.

Our finding may suggest that (i)
Ly$\alpha$ emitters at $z=6.56$ are
brighter intrinsically than those at $z=5.7$, or (ii)
the intrinsic luminosity is the same, but 
more
Ly$\alpha$ photons escaped from galaxies at $z=6.56$ than from $z=5.7$.
(The absorption in the IGM was kept the same.) 
The possibility (ii) is quite plausible, if dust content
of galaxies at $z=6.56$ is less than that at $z=5.7$.
How much less requires a more careful analysis,
which we shall give in section~\ref{sec:anomaly}.

In order to put better constraints on the bright end of the luminosity function, a larger survey is needed.  The brightest Ly$\alpha$ emitter detected at z=5.7 had a narrow band magnitude of 23.41 and the brightest at z=6.56 had a narrow band magnitude of 24.13.  To detect brighter Ly$\alpha$ emitters, a larger survey would be needed that could find the rare, high density peaks of the mass function.  In order to detect 10 galaxies above a magnitude of 23.41 at z=5.7, the survey would need to be at least $6.34$ degree$^2$ if we assume $M_h/L_{band}$ of 120, and at least $3.13$ degree$^2$ if we assume a $M_h/L_{band}$ of 95.  For the Subaru field at z=6.56, to detect 10 galaxies with a narrow band magnitude of 24.13 or higher, the area of the survey would have to be $3.6$ degree$^2$ if we assume a $M_h/L_{band}$ of 100 and $2.2$ degree$^2$ if we assume a $M_h/L_{band}$ of 85.

The LALA data \citep{rhoads/etal:2004} also probe a very similar
redshift, $z=6.55$.
Since only one galaxy was found from LALA, the Poisson error is large.
Nevertheless, the LALA data give us an important cross-check of
the results obtained from the Subaru field at the same redshift.
We find that $M_h/L_{band}\sim 160$ explains LALA's
detection of one galaxy at $z=6.55$. When we compare the LALA and Subaru
counts,  
we must take into account the different bandwidths of
these surveys. The LALA's bandwidth is about 60\% narrower than 
Subaru's
(Table~\ref{table:surveyprop}), 
and thus the constraint from the LALA data would correspond to
$M_h/L_{band}\sim 100$ for the Subaru data. We thus conclude that 
the constraints from the Subaru and LALA fields at $z=6.55$ are
comfortably consistent with each other. A more thorough comparison will
be given in section~\ref{sec:results}.

At $z=7.0$ \citet{iye/etal:2006} discovered one Ly$\alpha$ emitter in
the Subaru field that
was conformed spectroscopically. While the Poisson error is large, we
find that $M_h/L_{band}\sim 100$ explains Subaru's detection of one
galaxy at $z=7.0$. This number is remarkably similar to what we have
found from the Subaru fields at $z=5.7$ and $6.56$ as well as from the
LALA field at $z=6.55$.

At $z=8.76$ the searches performed in the ISAAC/VLT fields yielded 
null results.
We therefore place lower limits to $M_h/L_{band}>7$ and 3 from
the ZEN and ISAAC ext, respectively.
The weaker constraint from the latter is due to a brighter
flux density limit. These lower limits are consistent with properties of Ly$\alpha$
emitters as constrained by the other searches at $z\le 7$.

In order to improve upon these results, how large of a survey would be needed?  In order for 10 galaxies above the flux density limit to be seen in the Subaru (z=7.025) and the LALA fields, the survey area would need to be increased to $2.5$ degree$^2$ and $3.4$ degree$^2$ respectively.  If we assume that the mass to light ratio and $\alpha_{esc}$ is the same at z=8.76 than it is at z=7.025, the survey area of the ZEN and ISAAC ext fields would need to be increased to $24.8$ and $1.69\times 10^5$ degree$^2$ respectively.  The area needed for the ISAAC ext field is larger than the entire sky, so it becomes apparent that an adjustment to the bandwidth or flux density detection limit is necessary to make finding a Ly$\alpha$ emitter more feasible.

In summary, properties of
Ly$\alpha$ emitters and the IGM opacity 
have not evolved very much between $z=5.7$ and 7.
The lack of detection at $z=8.76$ is also consistent with no evolution,
although it does not provide a significant constraint yet.

\subsection{Finding a mass-to-``bolometric light'' ratio}
\label{sec:lbol}
\begin{table*}
\begin{minipage}{14cm}
\caption{%
 Fitting functions for the number of hydrogen ionizing photons per
 second, $Q(H)$, 
 stellar temperature of the star, $T_{eff}$, bolometric luminosity of
 the star,
 $L^*_{bol}$, and stellar lifetime, $\tau_*$, for varying metallicities.  
 These
 were obtained from stellar models from \citet{marigo/etal:2001}$^a$,
 \citet{lejeune/schaerer:2001}$^b$, and \citet{schaerer:2002}$^c$, or
 the fitting functions from  \citet{schaerer:2002}$^d$. 
 Note that $y\equiv \log(M_*/M_{\sun})$, and $\log$ is a logarithm of base 10.
}%
\begin{center}
\begin{tabular}{|l||l|l|}
\hline
$Z = 0$ & Fitting Function  & Ref. \\
\hline
$\log[Q(H)/{\rm s}^{-1}]$  &  $43.61 + 4.90y-0.83y^2$ for $9\leq M_*\leq500{\rm M_{\sun}} $ & d \\
& $39.29+8.55y$ for $5\leq M_*<9 {\rm M_{\sun}}$ & d\\
& $0$ for $M_*<5$& \\
$\log[T_{eff}/{\rm K}]$  &  $3.64+1.50y-0.556y^2+0.070y^3$ for $M_*\geq10 {\rm M_{\sun}}$ & c\\
& $3.87+0.937y-0.156y^2 $ for  $M_*<10 {\rm M_{\sun}}$& a\\
$\log[L^*_{bol}/L_{\sun}]$  &  $0.457+3.90y-0.530y^2$ for $M_*\geq10 {\rm M_{\sun}}$&c\\
&$0.219+4.51y-0.923y^2$  for  $M_*<10 {\rm M_{\sun}}$& a\\
$\log[\tau_*/{\rm year}]$  &  $9.79-3.76y+1.41y^2-0.186y^3$ for $M_*\geq10 {\rm M_{\sun}}$&c\\
&$9.80-3.52y+1.18y^2-0.164y^3 $  for  $M_*<10 {\rm M_{\sun}}$& a\\
\hline
\hline
$Z = 1/50~Z_{\sun}$&  &\\
\hline
$\log[Q(H)/{\rm s}^{-1}]$   &  $ 27.80+30.68y-14.80y^2+2.50y^3 $ for $M_*\geq5$&d\\
& $0$ for $M_*<5$& \\
$\log[T_{eff}/{\rm K}]$  &  $3.86+0.898y-0.299y^2+0.0399y^3$&b\\
$\log[L^*_{bol}/L_{\sun}]$   &  $0.175+4.26y-0.652y^2$&b\\
$\log[\tau_*/{\rm year}]$  &  $9.82-3.13y+0.743y^2$&b\\
\hline
\hline
$Z = 1~Z_{\sun}$&  &\\
\hline
$\log[Q(H)/{\rm s}^{-1}]$   &  $27.89+27.75y-11.87y^2+1.73y^3$ for $M_*\geq5$&d\\
& $0$ for $M_*<5$& \\
$\log[T_{eff}/{\rm K}]$   &$3.74+0.826y-0.166y^2$ &b\\
$\log[L^*_{bol}/L_{\sun}]$  &  $-0.148+4.70y-0.781y^2$&b\\
$\log[\tau_*/{\rm year}]$  &  $10.08-3.47y+0.774y^2+0.0327y^3$&b\\
\hline
\end{tabular}
\label{table:fitfunc}
\end{center}
\end{minipage}
\end{table*}


\begin{figure*}
\begin{minipage}{14cm}
\centering \noindent
\includegraphics[width=13cm]{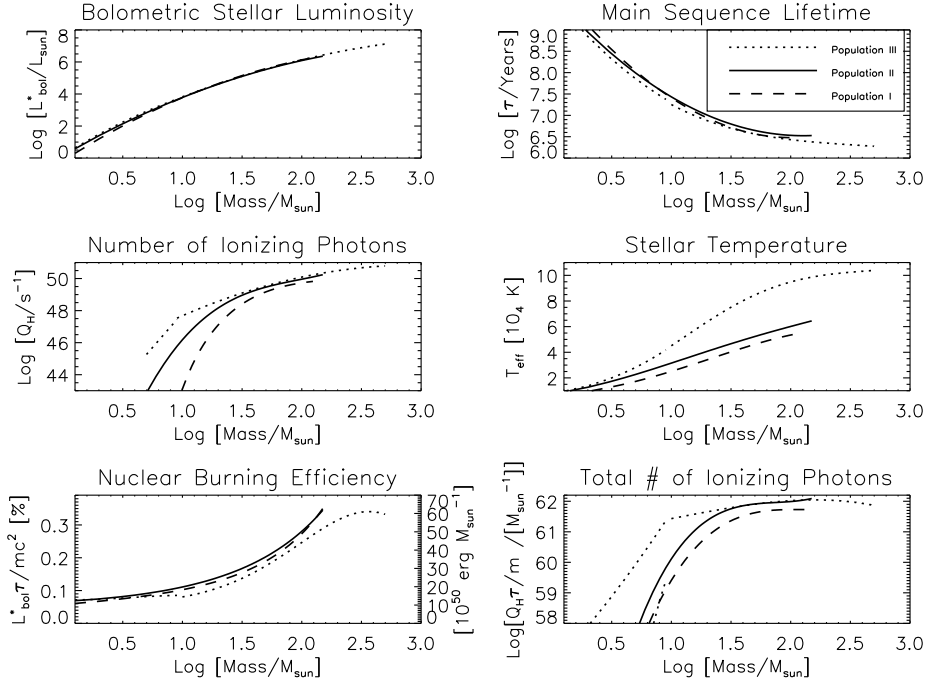}
\caption{%
 Fitting functions for the theoretical stellar data for the
 stellar bolometric luminosity, $L^*_{bol}$, lifetime, $\tau_*$, the number of
 hydrogen ionizing photons per second, $Q(H)$, and 
 the stellar temperature,  $T_{eff}$, for varying metallicities:
 $Z=0$ (Population III), $Z=1/50~Z_{\sun}$ (Population II), and
 $Z=1~Z_{\sun}$ (Population I). 
 The
 fitting formulae are given in Table~\ref{table:fitfunc}. 
 In addition, the nuclear
 burning efficiency (the total radiation energy produced over the
 stellar lifetime divided by the rest mass energy, $M_* c^2$), 
 and the total number of ionizing
 photons over the star's lifetime are plotted. 
}
\label{fig:fitfunc}
\end{minipage}
\end{figure*}

\begin{figure*}
\begin{minipage}{14cm}
\centering \noindent
\includegraphics[width=14cm]{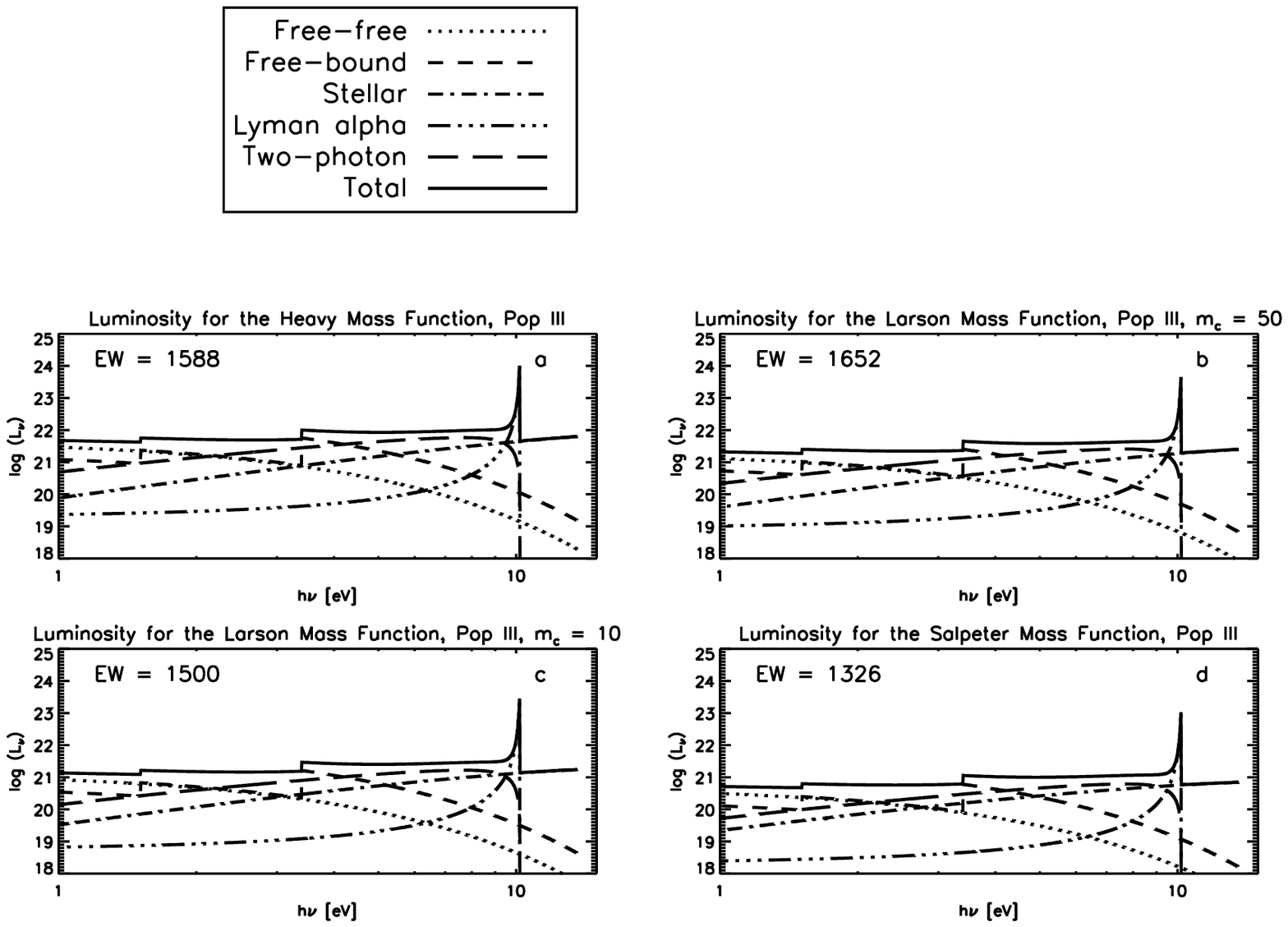}
\includegraphics[width=14cm]{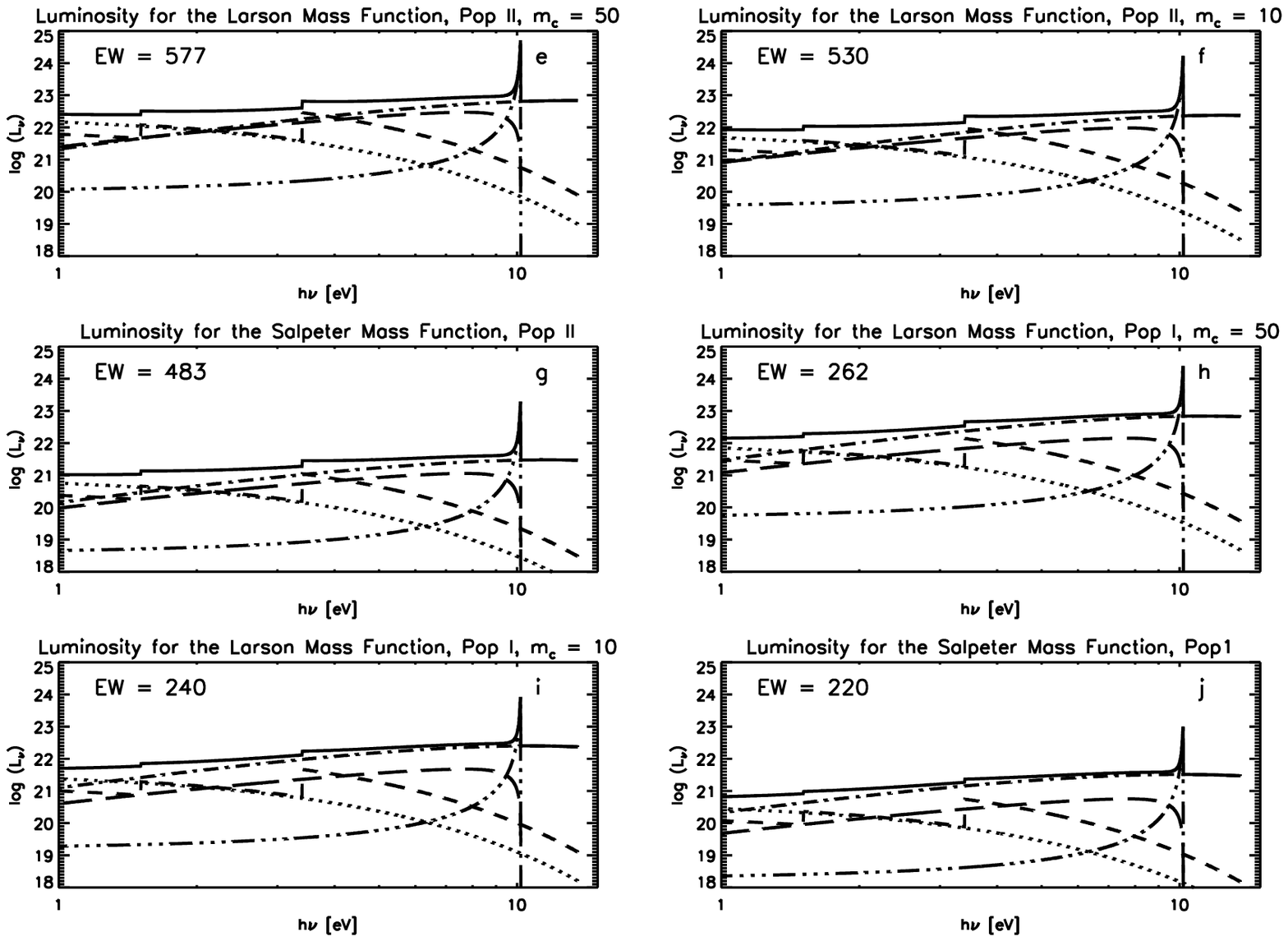}
\caption{%
Rest-frame spectra of galaxies with various populations of stars integrated
 over a mass spectrum. 
The vertical axis is in units of
 erg~s$^{-1}$~Hz$^{-1}$~M$_{\sun}^{-1}$.
The solid lines show the total spectra, while the
 dotted, short dashed, dot-dashed, dot-dot-dot-dashed, and long dashed
 lines show the free-free, free-bound, stellar, Ly$\alpha$, and
 two-photon emission, respectively. We have adopted a
 Ly$\alpha$ profile from
 \citet{loeb/rybicki:1999,santos/bromm/kamionkowski:2002}. 
 (For further discussion of the profile, see Appendix~\ref{sec:profile}.)
 We show the luminosity averaged over various
 mass spectra, given in section \ref{sec:interpretation}, divided by the
 mean stellar mass. 
 The spectra are computed for
 a galaxy at $z=7.025$, although the redshift affects the shape of the
 Ly$\alpha$ line profile  only.  The equivalent width (EW) of the Ly$\alpha$ 
line before extinction or scattering 
is also given. The EW has been computed from 
$\mbox{EW}=(\mbox{Total flux in Ly$\alpha$})/(\mbox{Continuum flux at
 1216~\AA})$. 
}%
\label{fig:spectra}
\end{minipage}
\end{figure*}

\begin{table*}
\begin{minipage}{16cm}
\caption{%
  Ratio of luminosity observed within the bandwidths of Subaru
 ($L_{Sub,z=5.7}$, Shimasaku et al. 2006; $L_{Sub,z=6.56}$, Taniguchi et
 al. 2005; Kashikawa et
 al. 2006; $L_{Sub,z=7}$, Iye et al. 2006), LALA ($L_{LALA}$,
 Rhoads et al. 2004), and ZEN and ISAAC ext 
 ($L_{ISAAC}$, Willis \& Courbin 2005; Willis et al. 2006; Cuby
 et al. 2007) in comparison to the bolometric luminosity,
 $L_{bol}$, for a variety of mass spectra and metallicities, in the
 absence of absorption or extinction of Ly$\alpha$. 
 The ratio
 is nearly independent of stellar mass spectra, while it drops as the
 metallicity of stars increases.   
}%
\begin{center}
\begin{tabular}{|l|l|l||l|l|l|l|l|}
\hline
Metallicity ($Z_{\sun}$) & $f(m)$ & $m_1, m_2 ({\rm M_{\sun}})$ & $\frac{L_{Sub_{z=5.7}}}{L_{bol}} $ &$\frac{L_{LALA}}{L_{bol}} $&$\frac{L_{Sub_{z=6.56}}}{L_{bol}} $  & $\frac{L_{Sub_{z=7}}}{L_{bol}}$ & $\frac{L_{ISAAC}}{L_{bol}} $   \\
\hline        
$0$ & $300 {\rm M_{\sun}} \delta$-function &-- &0.312 &0.209&0.263   & 0.297 &  0.120\\
$0$ & Heavy & 100, 500 &0.311 &0.209 &0.262&0.297& 0.120\\
 $0$ & Larson,$m_c = 50 {\rm M_{\sun}}$& 0.8, 150& 0.313&0.210&0.264 &0.299&0.120\\  
 $0$ & Larson,$m_c = 10 {\rm M_{\sun}}$& 0.8, 150& 0.305&0.205&0.257 &0.291&0.117\\
 $0$ & Salpeter & 0.8, 150 & 0.293&0.197&0.247 & 0.280& 0.113\\
$1/50$ & Larson, $m_c = 50 {\rm M_{\sun}}$  & 0.8, 150 & 0.219&0.146&0.185 &0.210 & 0.0843 \\
$1/50$ & Larson, $m_c = 10 {\rm M_{\sun}}$ & 0.8, 150 & 0.209&0.139&0.176 &0.201& 0.0806 \\
$1/50$ & Salpeter & 0.8, 150 &0.198&0.131 & 0.167&0.190& 0.0761\\
$1$  & Larson, $m_c = 50 {\rm M_{\sun}}$  & 0.8, 120 & 0.143&0.0945&0.121 &0.138 & 0.0555 \\
$1$ & Larson, $m_c = 10 {\rm M_{\sun}}$ & 0.8, 120 &0.134 &0.0881&0.113 &0.130& 0.0519 \\
$1$ & Salpeter & 0.8, 120 & 0.124&0.0813 &0.105&0.120& 0.0480\\
\hline
\end{tabular}
\label{tab:Lumratio}
\end{center}
\end{minipage}
\end{table*}


Our analysis so far has been relatively model-independent.
We have extracted the only free parameter, $M_h/L_{band}$, from various
narrow-band searches of Ly$\alpha$ emitters.

Here, $M_h/L_{band}$ only describes the light observed over the narrow
band (the luminosity within the bandwidth, $L_{band}$, not the
bolometric luminosity, $L_{bol}$).  
To proceed further and understand physical properties of
Ly$\alpha$ emitters better, however, we must relate $M_h/L_{band}$ to
the mass-to-``bolometric light'' ratio, $M_h/L_{bol}$, taking into
account stellar populations as well as differences in the bandwidths.

In order to get the actual mass to light ratio,
the spectra of a stellar population of galaxies must be modeled and
integrated first over all frequencies and then compared to the light
that is observed in the narrow band.  The fraction of Ly$\alpha$ photons
that survived, $\alpha_{esc}$, also needs to be taken into account.  

As a result, each data-set yields a constraint on 
$(M_h/L_{bol})\alpha_{esc}^{-1}$ as a function of 
assumed stellar populations. 

When the duty cycle is less than unity, the
constraint should be interpreted as
$(M_h/L_{bol})(\alpha_{esc}\epsilon^{1/\gamma})^{-1}$, where
 $\gamma\sim 2$ is a local slope of the cumulative luminosity function,
 $N(>L)\propto L^{-\gamma}$, to which  the current data are sensitive.

The spectra of a population of stars depend on the stellar mass spectrum
and 
metallicity of stars.  
We shall use a variety of mass functions paired with
metallicities: 
(a) Salpeter \citep{salpeter:1955}:
\begin{equation}
  f(m) \propto m^{-2.35} ,
  \label{eq:salpeter}
\end{equation}
(b) Larson \citep{larson:1998}:
\begin{equation}
  f(m) \propto m^{-1}\left(1+\frac{m}{m_c}\right)^{-1.35} ,
  \label{eq:larson}
\end{equation}
which matches Salpeter's in the limit of 
$m_c\rightarrow 0$.  One can explore a variety of models
by changing one parameter, $m_c$. 
(c) A top-heavy spectrum:
\begin{equation}
  f(m) \propto \left\{ 
  \begin{array}{ll}
  m^{-1}, & 100<m<500~{\rm M_\odot}\\
  0, & {\rm otherwise} ,
  \end{array}\right.
  \label{eq:topheavy}
\end{equation}
which might be possible 
for the primordial metal-free stars \citep{bromm/larson:2004}. 
(Note that $mf(m)$ is flat for $100<m<500~{\rm M_\odot}$.)
The normalizations are given by 
\begin{equation}
  \int_{m_1}^{m_2} dm~f(m) = 1 ,
\end{equation}
with $m_1$ and $m_2$ being the mass limits that the mass function is integrated over.  We also consider a delta function mass spectrum, with populations consisting of only $300~{\rm M_{\sun}}$ stars.

The synthetic spectrum emerging from a galaxy with a given population of
stars is the result of a variety of radiation
processes.  Some of the light from the star is converted by the nebula
into the Ly$\alpha$ line, free-free, free-bound, and two-photon
emission.  

We use analytical formulae for these spectra given in
section 2 of \citet{fernandez/komatsu:2006}, paired with a line profile
of Ly$\alpha$ emission
from \citet{loeb/rybicki:1999,santos/bromm/kamionkowski:2002}.  
These formulae are fully analytic and therefore it is easy to adjust
various parameters. 
We have checked that our predicted luminosities agree well with those produced
from the numerical code, {\sc cloudy}.  

Care must be taken when one computes the profile of Ly$\alpha$ line.
The width of the line profile we adopt here is likely too
broad, as this profile assumes that the IGM around a source is
completely neutral.
The other extreme case, a delta function profile at 10.2~eV,
increases the inferred $M_h/L_{bol}$ by a factor of at most a few.
We study this issue further in Appendix~\ref{sec:profile}.

We explore three metallicities:
what we refer to as Population III ($Z=0$), Population II
($Z=1/50~Z_{\sun}$), and Population I ($Z=1~Z_{\sun}$).  For convenience
we have fit the theoretical stellar data for bolometric luminosity,
stellar temperature, lifetime, and number of ionizing photons.  The
fitting functions, given in Table \ref{table:fitfunc} and plotted in
Fig. \ref{fig:fitfunc}, have been obtained by fitting stellar models
provided by the papers given in the third column of Table
\ref{table:fitfunc}. 

In Fig. \ref{fig:spectra}, we show the luminosity integrated over a mass
spectrum divided by the average stellar mass:   
\begin{equation}
\frac{\int^{m_2}_{m_1} L_{\nu,i} f(m) dm}{\int^{m_2}_{m_1} m f(m) dm}
\end{equation}
for each component $i$, which includes stellar black-body, Ly$\alpha$,
free-free, free-bound, and two-photon.   

Now we are in a position to calculate a conversion factor from
$L_{band}$ to $L_{bol}$: 
\begin{equation}
   \frac{L_{band}}{L_{bol}}= \frac
    {\int^{\nu_{max}}_{\nu_{min}}d\nu\sum_i \int^{m_2}_{m_1} L_{\nu,i}
    f(m) dm}{\int^{\infty}_{0}d\nu \sum_i \int^{m_2}_{m_1} L_{\nu,i}
    f(m) dm}, 
\end{equation}
where $\nu_{max}=\nu_{2,obs}(1+z)$ 
and $\nu_{min}=\nu_{1,obs}(1+z)$ are the limiting frequencies of 
instruments in the rest frame of the galaxies (again assuming a rectangular bandpass), $L_{\nu,i}$ is the luminosity of 
each component, and
$m_2$ and $m_1$ are the mass limits of the stellar mass spectrum, $f(m)$.

In Table \ref{tab:Lumratio} we show $L_{band}/L_{bol}$ for the various
surveys.  We find that  $L_{band}/L_{bol}$ is fairly constant over
different mass spectra, but depends mainly on metallicity.  
Metal-free stars sustain
higher temperatures as they undergo nuclear burning through the p-p
chain.  Because of this, their stellar spectrum is harder than stars with
metals (see panels a-d in Fig.~\ref{fig:spectra}).  Therefore they emit more ionizing photons that can be converted
by the surrounding nebula into the Ly$\alpha$ line.
As the metallicity
increases (panels e-j), the stellar temperature decreases, the stellar spectrum softens, the ionizing
photon flux decreases, and thus 
the Ly$\alpha$ line is depleted for a given stellar mass.
As a result, one obtains  a lower $L_{band}/L_{bol}$ for a higher
metallicity.

Now, calculating the mass-to-bolometric light ratio of galaxies is simple:
multiply the mass-to-observed light ratio ($M_h/L_{band}$) by
the ratio of observed to bolometric luminosity given in Table
\ref{tab:Lumratio}. 

This is, however, not the end of story. 
Not all Ly$\alpha$ photons would
escape from galaxies due to dust extinction, or from the IGM due to
resonant scattering. We are unable to distinguish these two effects;
thus, we parametrize a combined effect by a single parameter,
$\alpha_{esc}$, a survival fraction of Ly$\alpha$ photons.

\subsection{Results}
\label{sec:results}
\begin{table*}
\begin{minipage}{14cm}
\caption{%
 The mass (total halo mass) to light (bolometric
 luminosity) ratio times $1/(\alpha_{esc}~\epsilon^{1/\gamma})$.  
 The luminosity refers to the
 intrinsic luminosity before absorption or extinction of Ly$\alpha$
 photons. For each metallicity of stellar populations ($Z=0$,
 $Z=1/50~Z_{\sun}$, and $Z=1~Z_{\sun}$)
 a range of values represent a range of
 stellar mass spectra. For the observational data 
 we used $M_h/L_{band}=95-120$ and $85-100$ for the Subaru fields at $z=5.7$ and
 6.56, respectively, whereas we used $M_h/L_{band} \sim 160$ and $\sim 100$
 for the LALA field at $z=6.55$ and the Subaru field at $z=7.025$. 
 The latter values are much more uncertain than the former ones
 due to a large Poisson error,
 as only one Ly$\alpha$  emitter was found in each of the latter fields.
 For the ZEN and ISAAC fields only lower limits are given, as no sources
 were found in these fields. Note that $\gamma\sim 2$ for the surveys
 listed here.
}%
\begin{center}
\begin{tabular}{|l|l|l|l|l|}
\hline
Field & Redshift & $\frac{M_h}{L_{bol}}\frac{1}{\alpha_{esc}~\epsilon^{1/\gamma}}$ ($Z=0$)& $\frac{M_h}{L_{bol}}\frac{1}{\alpha_{esc}~\epsilon^{1/\gamma}}$ ($Z=1/50~Z_{\sun}$) &$\frac{M_h}{L_{bol}}\frac{1}{\alpha_{esc}~\epsilon^{1/\gamma}}$ ($Z=1~Z_{\sun}$) \\
\hline
Subaru&$ 5.7$ & $28-38$ & $19-26$ & $12 -17 $\\
LALA&$ 6.55 $ & $ \sim 32-34$ & $ \sim 21-23$ & $ \sim 13 -15 $\\
Subaru&$ 6.56$ & $21-26$ & $14 -19 $ & $8.9-12 $\\
Subaru & $ 7.025$ & $ \sim 28-30$&$ \sim 19 -21  $& $ \sim 12 -14 $  \\
ZEN &$ 8.76$ & $> 0.79-0.84$ & $> 0.53-0.59$ & $>0.34-0.39$\\
ISAAC ext &$ 8.76 $ & $> 0.34-0.36 $&$> 0.23-0.25$&$>0.14 -0.17 $ \\
\hline
\end{tabular}
\label{tab:masslight}
\end{center}
\end{minipage}
\end{table*}

The outcome of our analysis is a mass-to-bolometric light ratio
divided by a Ly$\alpha$ survival fraction and 
the effect of duty cycle, 
$(M_h/L_{bol})(\alpha_{esc}\epsilon^{1/\gamma})^{-1}$, where $L_{bol}$ is the {\it intrinsic}
luminosity of galaxies before absorption or extinction of Ly$\alpha$ photons.
We tabulate this quantity inferred from various
narrow-band searches in Table \ref{tab:masslight},
which is the main result of this paper.

Since the
effect of bandwidths has been taken into account properly, 
these constraints can be compared with each other on equal footing.
Let us analyze the results in Table \ref{tab:masslight}.   
As noted in the previous section, 
the Ly$\alpha$ line diminishes in strength and 
$(M_h/L_{bol})(\alpha_{esc}\epsilon^{1/\gamma})^{-1}$
drops as metallicity increases.  
A variation due to different stellar mass spectra is negligible.

The values of
$(M_h/L_{bol})(\alpha_{esc}\epsilon^{1/\gamma})^{-1}$ inferred from the current data 
processed through our
simple model are rather reasonable:
for all cases where at least one source is found per field,
the inferred $(M_h/L_{bol})(\alpha_{esc}\epsilon^{1/\gamma})^{-1}$ falls between 9 and 38,
the low and high values being for the solar and zero metallicity, respectively.
Consistency across redshifts ($z=5.7$, 6.5, and 7.0) as well as across
different observations is striking.

We conclude from these results that the Ly$\alpha$ emitters detected in 
these narrow-band surveys are
either normal galaxy populations with 
$M_h/L_{bol}\sim 10$ and having a fair fraction of Ly$\alpha$ photons escape,
$\alpha_{esc}\epsilon^{1/\gamma}\sim 0.5-1$, 
or starburst galaxies with
$M_h/L_{bol}\sim 0.1-1$ and
a smaller fraction of the Ly$\alpha$ photons escaped from the galaxies
themselves  {\it and} the surrounding IGM,
$\alpha_{esc}\epsilon^{1/\gamma} 
\sim 0.01-0.05$. 
Note that the degeneracy still allows for a possibility of having a
significant survival fraction from 
these starburst populations, e.g., 
$\alpha_{esc}\sim 0.5$, if $\epsilon^{1/\gamma}\sim 0.1$
(or $\epsilon\sim 0.01$ and $\gamma\sim 2$).

It is clear that ${\alpha_{esc}}$, $\epsilon$, and $M_h/L_{bol}$ are
completely degenerate: we can only constrain the product of these three
properties, not the individual properties. 

The ratio of the Ly$\alpha$ flux to the continuum flux helps to lift
this degeneracy partially. The observed equivalent width (EW) of Ly$\alpha$
emitters is on the order of 100~\AA~or larger
\citep[e.g.,][]{kashikawa/etal:2006}. 
In Figure~\ref{fig:spectra} we show the predicted EW,
\begin{equation}
 \mbox{EW} = \frac{\mbox{Total Flux in Ly$\alpha$}}{\mbox{Continuum Flux
  at 1216~\AA}},
\end{equation} 
Using the zero-age main sequence values for the luminosity (given in Table \ref{table:fitfunc}), we obtain equivalent widths of $1300-1700$ for $Z=0$,  $480-580$ for $Z=1/50~Z_{\sun}$, and  $220-260$ for $Z=1~Z_{\sun}$.  Therefore, a low survival fraction, 
$\alpha_{esc}\sim 0.1$, is required for low metallicity populations, 
while a high $\alpha_{esc}\sim 0.5$ is required for high metallicity
ones, in order to fit the observed EW.  However, as the age of stars within the galaxy increase, the fraction of ionizing photons to non-ionizing photons decrease, and thus less photons are converted into Ly$\alpha$ photons.  Therefore, the equivalent width may decrease with time - depending on the age of the galaxy and the rate of star formation.  \citep{charlot/fall:1993, leitherer/etal:1999, kudritzki/etal:2000, malhotra/rhoads:2002, schaerer:2003}

Assuming the zero-age main sequence luminosity of the stars,
there are two solutions left for $\epsilon\sim 1$: (i) Ly$\alpha$ emitters at $z\ge 5.7$ are normal
populations with $Z>1/50~Z_{\sun}$ and $\alpha_{esc}> 0.5$, or (ii)
they are starburst populations with $Z<1/50~Z_{\sun}$ and
$\alpha_{esc}<0.1$.  As stars age, the EW of the population will also decrease, allowing for larger values of $\alpha_{esc}$ for a $Z<1/50~Z_{\sun}$ population.
For $\epsilon<1$ other solutions are still allowed.

Having the Ly$\alpha$ line be diminished in flux by about an order of
magnitude is not a surprising effect.
Both the IGM and galaxies themselves are expected to scatter or absorb
Ly$\alpha$ photons efficiently.
\citet{dijkstra/etal:2006a} claim that the asymmetry 
in Ly$\alpha$ lines that has been seen in the current data already suggests
that the IGM only transmitted $10-30$\% of the Ly$\alpha$ flux.
Several authors 
 model the effect of dust 
\citep{hansen/oh:2006, verhamme/etal:2006}
and neutral
hydrogen 
\citep{laursen/sommer-larsen:2007}
within galaxies on Ly$\alpha$ photons, and find that
even a small amount of dust can easily
absorb Ly$\alpha$ photons, and the resulting line profiles may be
complex
due to a structure in the distribution of dust and outflows.
In addition, high opacity near the line
centre decreases the flux at the centre of the line.  However, if the
medium is clumpy, the ratio of Ly$\alpha$ to continuum photons might
actually be increased \citep{neufeld:1991, hansen/oh:2006}.
\citet{mcquinn/etal:2006} also study the effect of neutral hydrogen
on the Ly$\alpha$ line using cosmological simulations.
Once the universe is almost totally reionized,
there will not be much suppression of the Ly$\alpha$ line, but before
then, Ly$\alpha$ luminosity might be able to help probe the size of HII
bubbles -- the larger the bubble, the less suppression of the Ly$\alpha$
line \citep{haiman/cen:2005}.

The physics of this problem is complex; however, our results are
consistent with a depletion of the Ly$\alpha$ flux if indeed these
Ly$\alpha$ emitters  are starburst galaxies with $M_h/L_{bol}\sim 0.1-1$.


\subsection{Interesting features}
\label{sec:anomaly}
Is there any ``anomaly''? 
Let us focus on  the Subaru fields at $z=5.7$ and 6.56, as 
these are the most accurate data-sets.
We observe nearly 20--30\% decrease in $(M_h/L_{bol})(\alpha_{esc}\epsilon^{1/\gamma})^{-1}$
from $z=5.7$ to $z=6.56$.
We saw this trend in the preliminary analysis based upon
$M_h/L_{band}$  in section~\ref{sec:lband}. 
After a more careful analysis we still observe the same trend. 

Although subtle,
if this is indeed a real effect, what would be the implication?
This effect {\it cannot} be explained by having a smaller $\alpha_{esc}$
(hence a larger opacity for Ly$\alpha$ photons) at higher $z$.
Therefore, it is inconsistent with neutral
fraction in the IGM around sources being higher at higher $z$.
On the contrary, one needs to have a larger  $\alpha_{esc}$ -- hence a
smaller opacity for Ly$\alpha$ photons -- at higher $z$, perhaps due to
less dust content \citep{haiman/spaans:1999}.
An alternative possibility is that $M_h/L_{bol}$ was lower in the past,
i.e., the Ly$\alpha$ emitters were
intrinsically brighter at higher $z$, perhaps due to a more intense
starburst. Such a burst would create a large HII bubble around the
source, which also helps to increase $\alpha_{esc}$ by suppressing the IGM
opacity. It therefore seems easy to explain the 20--30\% decrease in
$(M_h/L_{bol})(\alpha_{esc}\epsilon^{1/\gamma})^{-1}$ from $z=5.7$ to $6.5$.
A similar trend has also been pointed out by \citet{stark/loeb/ellis:2007}.

As we show in Appendix~\ref{sec:profile}, the magnitude of this effect
is reduced to 10-20\% 
if we assume that a line profile of Ly$\alpha$
photons is a delta function at 10.2~eV. 

Another interesting feature in Table~\ref{tab:masslight}
is that $(M_h/L_{bol})(\alpha_{esc}\epsilon^{1/\gamma})^{-1}$ at $z=6.56$ for $Z=0$
agrees with that at $z=5.7$ for $Z=1/50~Z_{\sun}$,
and $(M_h/L_{bol})(\alpha_{esc}\epsilon^{1/\gamma})^{-1}$ at $z=6.56$ for $Z=1/50~Z_{\sun}$
agrees with that at $z=5.7$ for $Z=1~Z_{\sun}$.
While it seems a mere numerical coincidence, it might also be suggestive
of the metallicity evolution in Ly$\alpha$ emitters.

\section{Comparison with Previous Work}
\label{sec:compare}
A halo mass function as a tool for calculating the luminosity function
of Ly$\alpha$ emitters is not a new idea
\citep[e.g.,][]{haiman/spaans:1999,haiman/spaans/quataert:2000}. 

Novelty of our approach is the use of the mass-to-light ratio as a
fundamental parameter, which has a few advantages. 
In this section we make this point clear by 
comparing our results with recent work on a similar subject.

In addition, our analysis is new in that we have explored various
assumptions about metallicity and stellar mass spectra of
Ly$\alpha$ emitters.

\subsection{\citet{dijkstra/etal:2006b}}
 \citet{dijkstra/etal:2006b} computed $N(>F)$ by integrating the
halo mass function over mass above a certain flux density, $F$. In order to
relate the host halo mass to the observed luminosity, $L_\alpha$,
\footnote{Their luminosity, $L_\alpha$, 
is different from our $L_{band}$, as they
ignored the line profile, continuum, and bandwidth of instruments.
}
they used
\begin{equation}
 \frac{M_h}{L_\alpha}=0.128\times
\frac{t_{sys}/(100~{\rm Myr})}{\frac{\Omega_b}{\Omega_m}\eta\alpha_{esc}},
\label{eq:mldijk}
\end{equation}
which is their equation~(2) in our notation.
(Note that in our notation $M/L$ is always measured in units of 
$M_{\sun}L_{\sun}^{-1}$.)
Here, 
$\eta$ is the fraction of baryon mass
converted into stars, 
$t_{sys}=\epsilon t_{hub}$  is the duration of a starburst,
$t_{hub}$ is the Hubble time, and $\epsilon$ is the duty cycle.

In our approach $M_h/L_\alpha$ is the only free parameter, and the
effect of $\epsilon$ is included using the degeneracy line, 
$(M_h/L_\alpha)\epsilon^{-1/\gamma}=\mbox{constant}$.
Their approach was to divide $M_h/L_\alpha$ up further by introducing
two free parameters, $\epsilon$ and $\eta\alpha_{esc}$, and constrain 
these parameters simultaneously.
However, it is difficult to extract 
more than $M_h/L_\alpha$ from the observed luminosity function.
Figure~1 of \citet{dijkstra/etal:2006b} also shows that 
$\epsilon$ and $\eta\alpha_{esc}$ are strongly degenerate. 
In our opinion the current data do not allow for 
 two free parameters to be constrained well.
In addition,
the use of $M_h/L_\alpha$ as a parameter avoids the need to
specify the duration of a starburst or the fraction of baryon mass
converted into stars. 

They found that $\alpha_{esc}$ at $z=5.7$ inferred from \citet{shimasaku:2006}
and  $\alpha_{esc}$ at $z=6.5$ inferred from
\citet{taniguchi/etal:2005,kashikawa/etal:2006}  
are about the same, the ratio of the two being 
$\alpha_{esc,57}/\alpha_{esc,65}\sim 0.8-1.5$ for a prior on
$\epsilon$ of $\epsilon=0.5-0.03$.
We would find $\alpha_{esc,57}/\alpha_{esc,65}\sim 0.7-0.8$
(see Table~\ref{tab:masslight}), if we assumed that the intrinsic
properties of Ly$\alpha$ emitters did not change between these
redshifts. The other values of $\alpha_{esc}$ are permitted 
when we vary the intrinsic mass-to-light, $M_h/L_{bol}$, with
$(M_h/L_{bol})(\alpha_{esc}\epsilon^{1/\gamma})^{-1}$ held fixed. This
is essentially 
equivalent to their varying $\epsilon$ along the degeneracy line.

When an additional constraint from the luminosity function of UV
continuum was included in the analysis, 
they found that the constraints shifted slightly to
$\alpha_{esc,57}/\alpha_{esc,65}\sim 1.1-1.8$.
While we do not perform a joint analysis with the UV continuum
luminosity function in this paper, we would expect a similar shift in the
parameter constraint.

\subsection{\citet{salvaterra/ferrara:2006}}
\citet{salvaterra/ferrara:2006} used an stellar mass spectrum
of stars that is given by a delta function at $m_*=300~M_{\sun}$,
and related $M_h$ to $L_{band}$ as
\begin{equation}
 \frac{M_h}{L_{band}} = 
\frac{300~M_{\sun}}{{\int^{\nu_{max}}_{\nu_{min}}d\nu\sum_i
L_{\nu,i}(300~M_{\sun})}}
\frac1{\frac{\Omega_b}{\Omega_m}\eta},
\end{equation}
which can be obtained from their equation~(3), combined with
our equation~(\ref{eq:fluxlum}).
Their $M_h/L_{band}$ is therefore 
equal to about 10 times\footnote{$\Omega_m/(\Omega_b\eta)\sim 10$.}
the mass-to-light of a  metal-free star of $300~M_{\sun}$.

The lifetime of a starburst of their model galaxy is as short as the
lifetime of stars, which is only 2 Myr.
In other words, they assumed that these massive stars formed at once in
a galaxy, so
that the lifetime of starbursts was the shortest
possible time, equal to the lifetime of the star.
This creates a very short lived but extremely bright galaxy that could
easily be detected with current observations.  

The mass-to-bolometric light ratio
of their model galaxy was 
$M_h/L_{bol}=6.73\times 10^{-4}$  and 
$1.35 \times 10^{-3}$ 
for ``H-cooling'' and ``H$_2$-cooling''
haloes, respectively, assuming all Ly$\alpha$ photons escaped.
(They used $\eta=0.8$ and 0.4 
for H-cooling and H$_2$-cooling
haloes, respectively.)
These extreme values allow them to predict
that there should be
thousands of galaxies seen in the NICMOS Ultra Deep Field, where only three or
fewer were actually detected, and 400 to 700 in the ZEN field, where no
sources were detected.  
They also reported that almost all of the Spitzer counts should be attributed
to galaxies above $z\sim 8$. 

Their conclusion is driven by their fixed value of $M_h/L_{bol}$,
which seems rather low.
  Our formulation, which treats $M_h/L_{band}$ as a free parameter,
allows for dimmer galaxies by
spreading out the star formation over a much longer period than 
the stellar lifetime.
This allows us to obtain results that are consistent with observations. 
Note that their using a delta-function mass spectrum is not the source
of discrepancy. We can still fit the observations with a reasonable
$M_h/L_{bol}$ for the same mass spectrum.
The source of discrepancy is their assumption about an instantaneous
starburst in 2~Myr.

They used these bright galaxies
to fit the observed excess in the near infrared background. 
The main conclusion of \citet{salvaterra/ferrara:2006} is that
the excess near infrared background cannot be
mainly coming from high-$z$ galaxies at $z\ga 7$, as
they do not see these extremely bright galaxies in the NICMOS UDF, ZEN,
or Spitzer counts.

However, their argument does not rule out the high-$z$ galaxies being 
the origin of the 
near infrared background.
Using a simple argument based upon  energy conservation, 
we have shown in the previous paper \citep{fernandez/komatsu:2006} that 
the near infrared background measures only the total light
integrated over time, and thus one can obtain the same amount of near
infrared background by having either (i) extremely bright sources over an
extremely short time period, such as those invoked by
\citet{salvaterra/ferrara:2006}, or (ii) much dimmer sources over a
much longer time period. 
While \citet{salvaterra/ferrara:2006} have successfully shown that 
the first possibility is ruled out, they have not ruled out the second
possibility yet. 

\subsection{\citet{ledelliou/etal:2006}}
\citet{ledelliou/etal:2006} predict the luminosity functions of
Ly$\alpha$ emitters at redshifts from $3<z<6.6$, using cosmological
simulations coupled with a semi-analytical galaxy formation model.

Similar to ours and the other work, they assume that the
escape fraction of Ly$\alpha$ photons are independent of halo mass,
and find its value, $\alpha_{esc}=0.02$, by fitting
the observed luminosity function of Ly$\alpha$ emitters at $z\sim 3$.  
\citep[See][for a criticism on this
assumption.]{kobayashi/totani/nagashima:2007}

Since the halo mass function is also an essential ingredient in the
semi-analytical galaxy formation model, and they make the same
assumption about the escape fraction of Ly$\alpha$ photons, we expect
our predictions and theirs to agree well for the same set of parameters.
We find that we can fit the bright-end of their predicted
luminosity functions (their Fig.~1 for $z=7$) with
a population of starburst galaxies, $M_h/L_{bol}\sim 1$, which is a very
reasonable result.

We believe that our simple model captures the basic physics that goes into
their model, which is more sophisticated and complex.

At a fainter end, however, their luminosity function flattens out and
our calculations always over-predict the number of sources.
This is likely due to our assumption about a constant mass-to-light
ratio. It is expected that this assumption breaks down once a large mass
range is included in the analysis. The most economical way to improve
our model is to introduce a second free parameter, a slope of
mass-to-light, such that $L\propto M^\beta$, for instance.
As the observations improve in the future, a two-parameter model such as
this should be used.

\section{Conclusions}
\label{sec:conclusions}
A simple model based upon the halo mass function coupled with a constant
mass-to-light ratio fits the luminosity functions measured and
constrained by the
current generation of narrow-band Ly$\alpha$ surveys
from $5.7\le z\le 8.8$. We have explored various metallicities and
 stellar mass spectra.

The inferred mass-to-light ratios 
are consistent with no evolution in the properties of Ly$\alpha$
emitters or opacity in the IGM from $5.7\le z\le 7$.
 Therefore, the current data of the luminosity
functions do not provide evidence for 
the end of reionization. The data at $z=8.8$ do not yield a significant
constraint yet.

These mass-to-light ratios suggest that 
the Ly$\alpha$ emitters discovered in the current surveys are
either starburst galaxies with only a smaller fraction
of
Ly$\alpha$ 
photons escaped from galaxies themselves and the IGM, 
$\alpha_{esc}\epsilon^{1/\gamma}\sim 0.01-0.05$,
or normal populations with a fair fraction of
Ly$\alpha$ photons escaped,
$\alpha_{esc}\epsilon^{1/\gamma}\sim 0.5-1$.
The luminosity function alone cannot distinguish between these two
possibilities. 


For the duty cycle of order unity, $\epsilon\sim 1$,
the observed equivalent width of Ly$\alpha$ line indicates that
starburst populations are consistent with low metallicity populations with
$Z<1/50~Z_{\sun}$, while normal populations are consistent with high metallicity
populations. The other solutions are still allowed for $\epsilon<1$.
Note that a recent study of the SED of Ly$\alpha$ emitters
by \citet{nilsson/etal:2007} shows that the Ly$\alpha$
emitters at $z=3.15$ are consistent with a very low metallicity
population, $Z=1/200~Z_{\sun}$.

To constrain the properties of Ly$\alpha$ emitters further, one should
use asymmetric absorption features of the measured Ly$\alpha$ line
profiles to distinguish between them
\citep{miralda-escude:1998,miralda-escude/rees:1998,haiman:2002,santos:2004,tasitsiomi:2006}. The
best way to break degeneracy between $\epsilon$ and $\alpha_{esc}$ is to
detect the deviation of the cumulative luminosity function from a pure
power-law. In order to do this it is crucial to determine the bright end
of luminosity function more accurately.

We disagree with the conclusion reached by
\citet{salvaterra/ferrara:2006}
that no detection of Ly$\alpha$ emitters at $z=8.8$ excludes the excess near
infrared background being produced by galaxies at $z>7$. While they have
excluded the excess background coming from extremely bright starburst galaxies
with $M_h/L_{bol}\sim 10^{-3}$ and the lifetime of 2~Myr, their argument does
not exclude another possibility that the
excess background originates from galaxies with $M_h/L_{bol}\sim 0.1-1$ and the
lifetime comparable to the age of the universe at $z>7$. 
As the near infrared
background measures only the total amount of light integrated over time,
both scenarios result in the same amount of background light.
As we have shown in this paper, the latter scenario is consistent with
all the existing Ly$\alpha$ surveys from $5.7\le z\le8.8$.

There is a subtle hint that $20-30$\% more Ly$\alpha$
photons survived from $z=6.5$ than from $z=5.7$.
A number of factors need to be checked carefully before this conclusion is taken
seriously: the completeness correction and spectroscopic confirmation
rate of the observed luminosity function, the shape of Ly$\alpha$
line profiles (which is however not quite enough to make the effect go away; see
Appendix~\ref{sec:profile}), and accuracy of the evolution of the 
theoretical halo mass function in these redshifts. 
In addition, more elaborated theoretical models such as those 
described in Sec.~\ref{sec:justification} may be
necessary to test reality of this effect, while it is interesting that
the model with a duty cycle has also shown a similar trend
\citep{stark/loeb/ellis:2007}.

Our method should provide a simple tool for interpreting the galaxy number
count data in terms of the mass-to-light ratio. 
Or, for a given mass-to-light ratio as constrained by the existing data,
it can also be used to predict 
the luminosity functions, and thus it helps to design future Ly$\alpha$
surveys such as 
a follow-up survey at $z=7$ with Subaru, at $z=8.8$ with VLT, and 
a new survey at even higher redshifts with the James Webb Space
Telescope
\citep[see also][for an alternative way of making forecasts]{barton/etal:2004}.

\section*{Acknowledgments}
We would like to thank Kyungjin Ahn,  Volker Bromm, Neal Evans, Juna
Kollmeier, Paul Shapiro, Gregory Shields, and Chris Sneden for
discussions,
and Mark Dijkstra, Andrea Ferrara, and Tomonori Totani for their comments on the manuscript.
We would also like to thank Robert Kurucz for his help on the
compilation of the stellar data.
E.R.F.
acknowledges support from a Continuing Fellowship of the University of
Texas at Austin. E.K. acknowledges support from an Alfred P. Sloan Fellowship.

\appendix
\section{On the Profile of Ly$\alpha$ Lines}
\label{sec:profile}
Throughout this paper we have adopted a theoretical line profile
computed by \citet{loeb/rybicki:1999} and later fit by 
\citet{santos/bromm/kamionkowski:2002} (their equation~15).
This line profile is fairly broad (see dot-dot-dot-dashed lines in
Figure~\ref{fig:spectra}). The physical origin of this broadening 
is a combination of the IGM scattering and cosmological redshift.
Since their underlying assumption that the IGM around sources is neutral
may not be always valid, a care must be taken when one uses their
theoretical profiles. 
The broadening is reduced significantly when the IGM around sources
is ionized.

The shape of the line profile affects our analysis through the
bandwidths of instruments. 
Since we are dealing with narrow-band filters, instruments miss a large
fraction of Ly$\alpha$ photons if a line profile is broader than their
bandwidths. 
Therefore, if we assumed erroneously that the line profile was too
broad, then the inferred $M_h/L_{bol}$ would be too low. 

In order to quantify a possible uncertainty regarding the shape of
the line profile, we explore the extreme case where a line profile is a
delta function at 10.2~eV. Since the instruments would not miss any 
Ly$\alpha$ photons, the luminosity within the band, $L_{band}$, would 
increase.  

Table~\ref{tab:deltaratio} shows that a delta-function line profile
increases the luminosity within instrument's bandwidths substantially.
For metal-free stars with a heavy mass spectrum,
the luminosity increases by a factor of 2 for Subaru, 
a factor of 3 for LALA, and a factor of 5.5 for ISAAC.

In Table~\ref{tab:masslight2} we report the inferred
$(M_h/L_{bol})(\alpha_{esc}\epsilon^{1/\gamma})^{-1}$ from assuming a delta function line
profile. These values should be compared with those in
Table~\ref{tab:masslight}. While there are changes in the inferred 
$(M_h/L_{bol})(\alpha_{esc}\epsilon^{1/\gamma})^{-1}$ at the level of a
factor of a few, the main result from our analysis does not change:
the Ly$\alpha$ emitters discovered in these surveys are either 
starburst galaxies with $\alpha_{esc}\epsilon^{1/\gamma}\sim 0.01-0.1$, or 
normal galaxies with $\alpha_{esc}\epsilon^{1/\gamma}\sim 0.5-1$.
 
On the other hand, a hint that $(M_h/L_{bol})(\alpha_{esc}\epsilon^{1/\gamma})^{-1}$
at $z=6.56$ is smaller than that at $z=5.7$ is now less significant:
it's only a 10--20\% effect, rather than a 20--30\% effect.

\begin{table*}
\begin{minipage}{16cm}
\caption{%
The same as Table~\ref{tab:Lumratio}, but for a Ly$\alpha$ line profile
 being a delta function at 10.2~eV. Small differences between observations
 are due to differences in the continuum flux within bandwidths.
}%
\begin{center}
\begin{tabular}{|l|l|l||l|l|l|l|l|}
\hline
Metallicity ($Z_{\sun}$) & $f(m)$ & $m_1, m_2 ({\rm M_{\sun}})$ & $\frac{L_{Sub_{z=5.7}}}{L_{bol}} $ 
&$\frac{L_{LALA}}{L_{bol}} $&$\frac{L_{Sub_{z=6.56}}}{L_{bol}} $  & $\frac{L_{Sub_{z=7}}}{L_{bol}}$ &
 $\frac{L_{ISAAC}}{L_{bol}} $   \\
\hline        
$0$ & $300 {\rm M_{\sun}} \delta$-function &-- &0.667 &0.662&0.665   & 0.668 &  0.661\\
$0$ & Heavy & 100, 500 &0.666 &0.662 &0.665&0.668& 0.661\\
 $0$ & Larson,$m_c = 50 {\rm M_{\sun}}$& 0.8, 150& 0.673&0.668 & 0.671 & 0.674 & 0.667 \\
 $0$ & Larson,$m_c = 10 {\rm M_{\sun}}$& 0.8, 150& 0.644 & 0.639 & 0.643 & 0.646 & 0.638\\
 $0$ & Salpeter & 0.8, 150 & 0.603&0.598& 0.601 & 0.604 & 0.597\\
$1/50$ & Larson, $m_c = 50 {\rm M_{\sun}}$  & 0.8, 150 & 0.385&0.378 & 0.383 & 0.387 & 0.377\\
$1/50$ & Larson, $m_c = 10 {\rm M_{\sun}}$ & 0.8, 150 & 0.362&0.354& 0.359 & 0.363 & 0.353\\
$1/50$ & Salpeter & 0.8, 150 &0.326&0.334 & 0.331 & 0.336 & 0.325\\
$1$  & Larson, $m_c = 50 {\rm M_{\sun}}$  & 0.8, 120 & 0.218&0.209 & 0.215 & 0.220 & 0.207\\
$1$ & Larson, $m_c = 10 {\rm M_{\sun}}$ & 0.8, 120 &0.200 & 0.192 & 0.197& 0.202 & 0.190\\
$1$ & Salpeter & 0.8, 120 & 0.182& 0.174 & 0.179 & 0.184 & 0.172\\
\hline
\end{tabular}
\label{tab:deltaratio}
\end{center}
\end{minipage}
\end{table*}


\begin{table*}
\begin{minipage}{14cm}
\caption{%
  The same as Table~\ref{tab:masslight}, but for a Ly$\alpha$ line profile
 being a delta function at 10.2~eV.
}%
\begin{center}
\begin{tabular}{|l|l|l|l|l|}
\hline
Field & Redshift & $\frac{M_h}{L_{bol}}\frac{1}{\alpha_{esc}~\epsilon^{1/\gamma}}$ ($Z_{\sun}=0$)& $\frac{M_h}{L_{bol}}\frac{1}{\alpha_{esc}~\epsilon^{1/\gamma}}$ ($Z_{\sun}=1/50$) &$\frac{M_h}{L_{bol}}\frac{1}{\alpha_{esc}~\epsilon^{1/\gamma}}$ ($Z_{\sun}=1$) \\
\hline
Subaru&$ 5.7$ & $57-81$  & $31-46$ & $17-26$  \\
LALA&$ 6.55 $ & $\sim 96-107$ & $\sim 53-60$ &$\sim 28-33  $ \\
Subaru&$ 6.56$ & $51-67  $ & $28-38  $ & $15-22 $\\
Subaru & $ 7.025$ &$ \sim 60-67  $ &$\sim 34-39    $& $\sim 18-22    $  \\
ZEN &$ 8.76$ & $>4.2 -4.7 $ & $> 2.3 -2.6 $ & $> 1.2-1.4  $\\
ISAAC ext &$ 8.76 $ & $>1.8-2.0    $ &$> 0.98-1.1$&$> 0.52-0.62 $ \\
\hline
\end{tabular}
\label{tab:masslight2}
\end{center}
\end{minipage}
\end{table*}


\end{document}